\documentclass[manuscript]{aastex631}

\def\JB{{\rm Jy~beam^{-1}}}
\def\mJB{{\rm mJy~beam^{-1}}}
\def\uJB{{\rm \mu Jy~beam^{-1}}}

\def\kms{{\rm km~s^{-1}}}
\def\Msun{M_{\sun}}
\def\Lsun{L_{\sun}}
\def\Rsun{R_{\sun}}

\graphicspath{{./}{figures/}}

\begin{document}

\title{Early Planet Formation in Embedded Disks (eDisk) VI: \\ Kinematic Structures around the Very Low Mass Protostar IRAS 16253-2429}

\correspondingauthor{Yusuke Aso}
\email{yaso@kasi.re.kr}

\author[0000-0002-8238-7709]{Yusuke Aso}
\affiliation{Korea Astronomy and Space Science Institute, 776 Daedeok-daero, Yuseong-gu, Daejeon 34055, Republic of Korea}

\author[0000-0003-4022-4132]{Woojin Kwon}
\affiliation{Department of Earth Science Education, Seoul National University, 1 Gwanak-ro, Gwanak-gu, Seoul 08826, Republic of Korea}
\affiliation{SNU Astronomy Research Center, Seoul National University, 1 Gwanak-ro, Gwanak-gu, Seoul 08826, Republic of Korea}


\author[0000-0003-0998-5064]{Nagayoshi Ohashi}
\affil{Academia Sinica Institute of Astronomy and Astrophysics, 11F of Astronomy-Mathematics Building, AS/NTU, No.1, Sec. 4, Roosevelt Rd, Taipei 10617, Taiwan}

\author[0000-0001-9133-8047]{Jes K. J{\o}rgensen}
\affil{Niels Bohr Institute, University of Copenhagen, {\O}ster Voldgade 5--7, DK~1350 Copenhagen K., Denmark}

\author[0000-0002-6195-0152]{John J. Tobin}
\affil{National Radio Astronomy Observatory, 520 Edgemont Rd., Charlottesville, VA 22903 USA}
\author[0000-0003-3283-6884]{Yuri Aikawa}
\affiliation{Department of Astronomy, Graduate School of Science, The University of Tokyo, 7-3-1 Hongo, Bunkyo-ku, Tokyo 113-0033, Japan}

\author[0000-0003-4518-407X]{Itziar de Gregorio-Monsalvo}
\affiliation{European Southern Observatory, Alonso de Cordova 3107, Casilla 19, Vitacura, Santiago, Chile}

\author[0000-0002-9143-1433]{Ilseung Han}
\affiliation{Division of Astronomy and Space Science, University of Science and Technology, 217 Gajeong-ro, Yuseong-gu, Daejeon 34113, Republic of Korea}
\affiliation{Korea Astronomy and Space Science Institute, 776 Daedeok-daero, Yuseong-gu, Daejeon 34055, Republic of Korea}

\author[0000-0002-2902-4239]{Miyu Kido}
\affiliation{Department of Physics and Astronomy, Graduate School of Science and Engineering, Kagoshima University, 1-21-35 Korimoto, Kagoshima, Kagoshima 890-0065, Japan}

\author[0000-0003-2777-5861]{Patrick M. Koch}
\affil{Academia Sinica Institute of Astronomy and Astrophysics, 11F of Astronomy-Mathematics Building, AS/NTU, No.1, Sec. 4, Roosevelt Rd, Taipei 10617, Taiwan}

\author[0000-0001-5522-486X]{Shih-Ping Lai}
\affiliation{Institute of Astronomy, National Tsing Hua University, No. 101, Section 2, Kuang-Fu Road, Hsinchu 30013, Taiwan}
\affiliation{Center for Informatics and Computation in Astronomy, National Tsing Hua University, No. 101, Section 2, Kuang-Fu Road, Hsinchu 30013, Taiwan}
\affiliation{Department of Physics, National Tsing Hua University, No. 101, Section 2, Kuang-Fu Road, Hsinchu 30013, Taiwan}
\affil{Academia Sinica Institute of Astronomy and Astrophysics, 11F of Astronomy-Mathematics Building, AS/NTU, No.1, Sec. 4, Roosevelt Rd, Taipei 10617, Taiwan}

\author[0000-0002-3179-6334]{Chang Won Lee}
\affiliation{Division of Astronomy and Space Science, University of Science and Technology, 217 Gajeong-ro, Yuseong-gu, Daejeon 34113, Republic of Korea}
\affiliation{Korea Astronomy and Space Science Institute, 776 Daedeok-daero, Yuseong-gu, Daejeon 34055, Republic of Korea}

\author[0000-0003-3119-2087]{Jeong-Eun Lee}
\affiliation{Department of Physics and Astronomy, Seoul National University, 1 Gwanak-ro, Gwanak-gu, Seoul 08826, Republic of Korea}

\author[0000-0002-7402-6487]{Zhi-Yun Li}
\affiliation{University of Virginia, 530 McCormick Rd., Charlottesville, Virginia 22904, USA}

\author[0000-0001-7233-4171]{Zhe-Yu Daniel Lin}
\affiliation{University of Virginia, 530 McCormick Rd., Charlottesville, Virginia 22904, USA}

\author[0000-0002-4540-6587]{Leslie W. Looney}
\affiliation{Department of Astronomy, University of Illinois, 1002 West Green St, Urbana, IL 61801, USA}

\author[0000-0002-0244-6650]{Suchitra Narayanan}
\affiliation{Institute for Astronomy, University of Hawai`i at Mānoa, 2680 Woodlawn Dr., Honolulu, HI 96822, USA}

\author[0000-0002-4372-5509]{Nguyen Thi Phuong}
\affiliation{Korea Astronomy and Space Science Institute, 776 Daedeok-daero, Yuseong-gu, Daejeon 34055, Republic of Korea}
\affiliation{Department of Astrophysics, Vietnam National Space Center, Vietnam Academy of Science and Technology, 18 Hoang Quoc Viet, Cau Giay, Hanoi, Vietnam}

\author[0000-0003-4361-5577]{Jinshi Sai (Insa Choi)}
\affil{Academia Sinica Institute of Astronomy and Astrophysics, 11F of Astronomy-Mathematics Building, AS/NTU, No.1, Sec. 4, Roosevelt Rd, Taipei 10617, Taiwan}

\author[0000-0003-1549-6435]{Kazuya Saigo}
\affiliation{Department of Physics and Astronomy, Graduate School of Science and Engineering, Kagoshima University, 1-21-35 Korimoto, Kagoshima, Kagoshima 890-0065, Japan}

\author[0000-0001-6267-2820]{Alejandro Santamar\'{i}a-Miranda}
\affiliation{European Southern Observatory, Alonso de Cordova 3107, Casilla 19, Vitacura, Santiago, Chile}

\author[0000-0002-0549-544X]{Rajeeb Sharma}
\affil{Niels Bohr Institute, University of Copenhagen, {\O}ster Voldgade 5--7, DK~1350 Copenhagen K., Denmark}

\author[0000-0003-0845-128X]{Shigehisa Takakuwa}
\affiliation{Department of Physics and Astronomy, Graduate School of Science and Engineering, Kagoshima University, 1-21-35 Korimoto, Kagoshima, Kagoshima 890-0065, Japan}
\affil{Academia Sinica Institute of Astronomy and Astrophysics, 11F of Astronomy-Mathematics Building, AS/NTU, No.1, Sec. 4, Roosevelt Rd, Taipei 10617, Taiwan}

\author[0000-0003-0334-1583]{Travis J. Thieme}
\affiliation{Institute of Astronomy, National Tsing Hua University, No. 101, Section 2, Kuang-Fu Road, Hsinchu 30013, Taiwan}
\affiliation{Center for Informatics and Computation in Astronomy, National Tsing Hua University, No. 101, Section 2, Kuang-Fu Road, Hsinchu 30013, Taiwan}
\affiliation{Department of Physics, National Tsing Hua University, No. 101, Section 2, Kuang-Fu Road, Hsinchu 30013, Taiwan}

\author[0000-0001-8105-8113]{Kengo Tomida}
\affiliation{Astronomical Institute, Graduate School of Science, Tohoku University, Sendai 980-8578, Japan}

\author[0000-0001-5058-695X]{Jonathan P. Williams}
\affiliation{Institute for Astronomy, University of Hawai`i at M\={a}noa, 2680 Woodlawn Dr., Honolulu, HI 96822, USA}

\author[0000-0003-1412-893X]{Hsi-Wei Yen}
\affil{Academia Sinica Institute of Astronomy and Astrophysics, 11F of Astronomy-Mathematics Building, AS/NTU, No.1, Sec. 4, Roosevelt Rd, Taipei 10617, Taiwan}




\begin{abstract}
Precise estimates of protostellar masses are crucial to characterize the formation of stars of low masses down to brown-dwarfs (BDs; $M_* < 0.08~\Msun$). The most accurate estimation of protostellar mass uses the Keplerian rotation in the circumstellar disk around the protostar.
To apply the Keplerian rotation method to a protostar at the low-mass end, we have observed the Class 0 protostar IRAS 16253-2429 using the Atacama Large Millimeter/submillimeter Array (ALMA) in the 1.3 mm continuum at an angular resolution of $0\farcs 07$ (10~au), and in the $^{12}$CO, C$^{18}$O, $^{13}$CO ($J=2-1$), and SO ($J_N = 6_5-5_4$) molecular lines, as part of the ALMA Large Program Early Planet Formation in Embedded Disks (eDisk). The continuum emission traces a non-axisymmetric, disk-like structure perpendicular to the associated $^{12}$CO outflow. The position-velocity (PV) diagrams in the C$^{18}$O and $^{13}$CO lines can be interpreted as infalling and rotating motions. In contrast, the PV diagram along the major axis of the disk-like structure in the $^{12}$CO line allows us to identify Keplerian rotation. The central stellar mass and the disk radius are estimated to be $\sim $0.12-0.17~$\Msun$ and $\sim $13-19~au, respectively. The SO line suggests the existence of an accretion shock at a ring ($r\sim 28$~au) surrounding the disk and a streamer from the eastern side of the envelope. IRAS 16253-2429 is not a proto-BD but has a central stellar mass close to the BD mass regime, and our results provide a typical picture of such very low-mass protostars.

\end{abstract}

\keywords{Circumstellar disks (235) --- Protostars (1302) --- Low mass stars (2050)}

\section{Introduction} \label{sec:intro}
The low-mass end of star formation is connected to brown dwarf (BD) formation. Investigating such low-mass regimes is crucial to comprehensively understanding star formation. BDs are characterized by their masses that are not enough to fuse hydrogen ($M_* < 0.08~\Msun$) and are as numerous as hydrogen-burning stars \citep{chab02}. The formation process of BDs can be similar to or different from those of low-mass stars \citep{pa.no04}. Theoretical studies have suggested various mechanisms for BD formation, such as turbulent fragmentation of a molecular cloud \citep[e.g.,][]{bate12}, disk fragmentation \citep[e.g.,][]{st.wh09}, ejection from multiple young stellar systems \citep[e.g.,][]{ba.vo12}, photo-erosion of a prestellar core by OB stars \citep[e.g.,][]{wh.zi04}, and eroding outflows \citep[e.g.,][]{mach09}.
A point of view for testing these scenarios is whether or not physical quantities in star formation and BD formation follow the same scaling laws. \citet{kim19} reported that their sample of 15 proto-BD candidates have different scaling laws, than the laws among $\sim 60$ low-mass protostars, between the outflow force versus the luminosity and between the outflow force and the envelope mass. Protostars around the BD threshold i.e., proto-BDs or very low-mass protostars, are then important to determine down to which mass such a scaling law of protostars holds.
Recent observational studies have aimed to identify and characterize proto-BDs, as well as pre-BDs \citep{degre16, huel17, sant21}. To study star formation in this very low mass regime, it is, therefore, necessary to establish a method for precisely estimating the central mass of each protostar down to the mass regime around $M_* \sim 0.1~\Msun$.

The most direct method for estimating a central protostellar mass is to identify the Keplerian rotation in a circumstellar disk, as demonstrated in previous observations toward young stellar objects in the typical low mass regime \citep[$M_* \gtrsim 0.2~\Msun$;][]{yen17}.
In contrast, previous studies of two representative proto-BD candidates, IC348-SMM2E and L328-IRS, estimated masses in indirect methods, which do not verify that the rotational velocity has the radial profile of the Keplerian rotation but merely assume that the observed gas motion is the Keplerian rotation or use other kinematics.
The central mass of IC348-SMM2E is estimated to be $\sim 0.002-0.024~\Msun$ from its luminosity, mass accretion rate, and efficiencies of mass accretion \citep{pala14}. This range includes a dynamical mass of $\sim 16~M_{\rm Jup}$ estimated on the assumption that the velocity gradient in the C$^{18}$O $J=2-1$ line is due to the Keplerian rotation, based on Submillimeter Array (SMA) observations \citep{pala14}.
The central mass of L328-IRS is estimated to be $\sim 0.012-0.023~\Msun$ from its mass accretion rate converted from its outflow force \citep{lee18}.
The outflow force is estimated using ALMA observations in the $^{12}$CO $J=2-1$ line and converted to the mass accretion rate, assuming an entrainment efficiency of 0.25, a ratio of mass loss and accretion rates of 0.1, and a wind velocity of $15~\kms$.
Meanwhile, \citet{lee18} explained the C$^{18}$O and $^{13}$CO $J=2-1$ emission in L328-IRS by the Keplerian rotation and suggested a central mass, $\sim 0.27$-$0.30~\Msun$, significantly larger than the BD mass regime.
In addition to the disregard of Keplerian-disk identification, numerical simulations predict that the disk mass can be comparable to the central stellar mass during the early phase with $M_* \lesssim 0.1~\Msun$ \citep{mach10,mach14}. Such a massive disk may cause the rotational velocity to deviate from the Keplerian rotation determined only by $M_*$, preventing us from directly estimating $M_*$ in observations.
It is thus crucial to verify whether the direct method with the Keplerian disk identification can be applied for protostars down to the $M_*\sim 0.1$-$\Msun$ regime as is for the typical low mass protostars.

\subsection{Target IRAS 16253-2429}
The Class 0 protostar IRAS 16253-2429 (hereafter I16253) in the Ophiuchus star-forming region is a good target for the verification in that mass regime.
I16253 has a bolometric luminosity of $L_{\rm bol}=0.16~\Lsun$ and a bolometric temperature of $T_{\rm bol}=42$~K \citep{ohas23}. Its internal luminosity is estimated to be $L_{\rm int}\sim 0.08~\Lsun$ from the infrared luminosity of $L_{\rm IR}=0.046~\Lsun$ measured in $1.25-70~\micron$ and an empirical ratio of $L_{\rm int}/L_{\rm IR}\sim 1.7$ \citep{dunh08}.
This protostar is thus classified as a very low luminosity object (VeLLO), which suggests that this protostar may eventually evolve into a very-low mass star or a BD. 
We observed this protostar as a part of the ALMA large program ``Early Planet Formation in Embedded Disks" (eDisk) \citep[see the overview paper by][]{ohas23}. The main goal of the program is to reveal signs of planet formation in disks in the course of protostellar evolution. In addition to this main goal, I16253 was observed to investigate the protostellar evolution and the disk formation down to the BD mass regime.

The central stellar mass of this target in previous works has been estimated to be a wide range of $\sim 0.02$ to $\sim 0.12~\Msun$, depending on methods.
\citet{tobi12}, who observed I16253 in the N$_2$H$^+$ $J,F_1,F=1,2,2-0,1,1$ line using the Combined Array for Research in Millimeter-wave Astronomy (CARMA) at a resolution of $\sim 9\arcsec$, estimated the stellar mass to be $0.1~\Msun$ based on comparisons of a position-velocity (PV) diagram across the associated outflow on a 6000-au scale between the observations and a model of the rotating, collapsing envelope \citep[UCM model;][]{ulri76, ca.mo81}.
\citet{yen17} estimated the stellar mass to be $0.02^{+0.02}_{-0.01}~\Msun$ based on C$^{18}$O $J=2-1$ observations of the protostar using ALMA at an angular resolution of $\sim 1\arcsec$; they reproduced the observed PV diagram along the major and minor axes of the C$^{18}$O envelope with a model envelope having a rotation conserving its specific angular momentum and a radial free-fall motion.
\citet{hsie19} used a similar model of the envelope and estimated the stellar mass to be $\sim 0.028~\Msun$ by reproducing PV diagrams of $^{12}$CO and C$^{18}$O $J=2-1$ obtained with ALMA at an angular resolution of $0\farcs 1 - 0\farcs 4$.
Additionally, \citet{hsie19} also suggested the stellar mass of $0.12~\Msun$ with the assumption that the $^{12}$CO emission arises from a Keplerian disk; they found offsets of the $^{12}$CO emission, $\pm 7.8$~au away from the center, at two velocity channels of $V_{\rm LSR}-V_{\rm sys}=\pm 3.4~\kms$ by 2D Gaussian fitting. This mass derived on the assumption of Keplerian rotation can be different from the one derived on the assumption of free-fall motion, if the infall velocity is slower than the free-fall velocity, as reported in other protostars \citep{ohas14, aso15, aso17, sai22}. These previous works show controversy about the central stellar mass of I16253 around the BD mass regime (i.e., $M_* <0.08~\Msun$).


Previous observations using the Walraven Photometer estimated the distance of the Ophiuchus star-forming region to be $125\pm 25$~pc \citep{dege89}, while the same observation estimated the distance at a location of $(l, b)=(353.161\arcdeg, 15.936\arcdeg)$ closest to I16253 $(353.2\arcdeg, 16.5\arcdeg)$ to be 140~pc. This distance is consistent with an estimate using {\it Hipparcos} \citep[$120 - 160$~pc;][]{kn.ho98} and a new estimate by {\it Gaia} DR2, $139^{+9}_{-10}$~pc, at a location of $(353.2\arcdeg, 16.6\arcdeg)$ closest to I16253 \citep{zuck20}. In addition to the I16253 distance, the average distance of the whole Ophiuchus region was updated by {\it Gaia} DR2 to $144\pm 9$~pc \citep{zuck19}. We adopt the {\it Gaia} DR2 distance of 139~pc as the distance of I16253 in this paper.

\vskip\baselineskip

This work aims to estimate the central mass of the Class 0 protostar I16253 more accurately and precisely than the previous works and reveal the kinematic structures around the protostar. Our observations have $>2.5$ times better sensitivity in the CO isotopologue lines than that of previous observations in the same lines at similar angular and velocity resolutions to those of our observations. We aim to reveal kinematic structures, such as a rotating disk, an infalling envelope, and an outflow, using the molecular line emission detected at the higher signal to noise ratio.
The rest of the paper has the  following structure. Section~\ref{sec:obs} describes key points of the settings and data processing of our I16253 observations. Section~\ref{sec:res} shows the observed results in the 1.3-mm dust continuum and the $^{12}$CO, $^{13}$CO, C$^{18}$O, and SO lines. We analyze the asymmetry in the continuum image and radial profiles of the rotational velocity in Section~\ref{sec:ana}. The rotational velocity is found to follow the Keplerian rotation, which allows us to directly estimate the central mass of I16253. We discuss kinetic structures around I16253, showing a schematic picture (Figure \ref{fig:schem}) of those structures, in Section~\ref{sec:dis}. A summary of our results and interpretation is in Section~\ref{sec:conc}.

\section{Observations} \label{sec:obs}

The details of the observing strategy, spectral setups, and data reduction process are presented in \citet{ohas23}\footnote{The scripts used for reduction, including the self-calibration, can be found at \url{https://github.com/jjtobin/edisk} \citep{tobi23}.}. Here we briefly describe the key points specific to I16253
and summarize the observational parameters in Table \ref{tab:obs}.
We observed I16253 using ALMA in Cycle 8 with the antenna configuration of C-8 on 2021 October 5, 26, 27, and 28 (2019.1.00261.L). The total observing time with C-8 is $\sim 310$ min (5.2 hr), while the on-source observing time is $\sim 60$ min. The number of antennas with C-8 was 46, and the projected baseline length ranges from 52 to 10540~m. Any Gaussian component with FWHM$\gtrsim 2.8\arcsec$ ($\sim 400$~au) was resolved out by $\gtrsim 63\%$ with this shortest baseline length \citep{wi.we94}. The phase center is $(\alpha _{\rm ICRS}, \delta _{\rm ICRS})=(16^{\rm h}28^{\rm m}$21\fs 6, $-24^{\circ}36\arcmin 23\farcs 4)$.

To recover the more extended structure, we also observed I16253 using ALMA in Cycle 8 with the antenna configuration of C-5 on 2022 June 14 and 15 (2019.A.00034.S). The total observing time with C-5 is $\sim 170$ min (2.8 hr), while the on-source observing time is $\sim 30$ min. The number of antennas with C-5 was 43, and the projected baseline length ranges from 14 to 1290~m. The resolving out scale is $10\arcsec$ ($\sim 1400$~au) with this shortest baseline length. The phase center is the same as with C-8.

The ALMA observations were set up to cover spectral windows including CO isotopologues ($J=2-1$), SO ($J_N = 6_5 - 5_4$), and other molecular lines at Band 6. The spectral window for the $^{12}$CO ($J=2-1$) line has 3840 channels covering a 940 MHz bandwidth at an original frequency resolution of 224 kHz. Spectral windows for the $^{13}$CO, C$^{18}$O ($J=2-1$), SO ($J_N=6_5-5_4$), and H$_2$CO ($3_{21}-2_{20}$) lines have 960 channels covering a 59 MHz bandwidth at an original frequency resolution of 61 kHz. Spectral windows for the other lines have 3840 channels covering a 1.9 GHz bandwidth at an original frequency resolution of 488 kHz. Channels were binned to produce the velocity resolutions of 0.32, 0.17, 0.17, $0.17~\kms$, respectively, to make maps in the $^{13}$CO, C$^{18}$O, SO, and H$_2$CO lines. The velocity resolution for other lines is $1.34~\kms$ (Appendix \ref{sec:otherlines}). Continuum maps were made using line-free channels of the spectral windows including two wide spectral windows with a $\sim 1.8$~GHz bandwidth centering at $\sim 220$ to $\sim 230$~GHz. The absolute flux density accuracy of ALMA is $\sim 10\%$ in this frequency band.

All the imaging procedure was carried out with Common Astronomical Software Applications (CASA) \citep{mcmu07} version 6.2.1. The visibilities were Fourier transformed and cleaned with Briggs weighting and a robust parameter of $-2.0$, 0.0, 0.5, or 2.0, using the CASA task {\it tclean} at a pixel size of $0\farcs 02$. Continuum images adopt robust$=-2.0$, 0.0, and 2.0 to show the most compact, intermediate, and most extended components, respectively. The line images adopt robust$=0.5$ and 2.0 to show compact components and extended components, respectively. The line images do not have signal-to-noise (S/N) ratios high enough for robust$=-2$. The continuum images with robust$=-2$ and 2 are produced with tapering parameters of 3 and 1~M$\lambda$ ($\sim 0\farcs06$ and $\sim 0\farcs 18$), respectively. The tapering parameter of 3~M$\lambda$ was selected to focus more on extended components with robust$=2$, while the tapering parameter of 1~M$\lambda$ was selected to increase the S/N ratio of the image with robust$=-2$. All the line images are produced with a tapering parameter of 2~M$\lambda$ ($\sim 0\farcs 09$) to increase the S/N ratio.
The resultant angular resolution is $0\farcs 04-0\farcs 29$ for the continuum emission and $0\farcs 17 - 0\farcs 38$ for the line emission.
We also performed self-calibration for the continuum data using tasks in CASA ({\it tclean}, {\it gaincal}, and {\it applycal}).
While all the maps are primary-beam corrected in this paper with the primary beam size of $26\arcsec$, the root-mean-square noise levels of the line maps were measured in emission-free channels/areas before the primary-beam correction, which indicates the sensitivity achieved toward the phase center. 

\begin{deluxetable}{cccccc}
\tabletypesize{\footnotesize}
\tablecaption{Summary of the parameters of our ALMA observations toward the Class 0 protostar I16253. \label{tab:obs}}
\tablehead{
\multicolumn{2}{c}{Date}&\multicolumn{2}{c}{2021 October 5, 26, 27, \& 28}&\multicolumn{2}{c}{2022 June 14 \& 15}\\
\multicolumn{2}{c}{Projected baseline range}&\multicolumn{2}{c}{52--10540 m}&\multicolumn{2}{c}{14--1290 m}\\
\multicolumn{2}{c}{Maximum Recoverable Scale}&\multicolumn{2}{c}{${0\farcs 62}$}&\multicolumn{2}{c}{${2\farcs 9}$}\\
\multicolumn{2}{c}{Bandpass/flux calibrator}&\multicolumn{2}{c}{J1517-2422}&\multicolumn{2}{c}{J1517-2422}\\
\multicolumn{2}{c}{Check source}&\multicolumn{2}{c}{J1650-2943}&\multicolumn{2}{c}{J1650-2943}\\
\multicolumn{2}{c}{Phase calibrator}&\multicolumn{2}{c}{J1633-2557}&\multicolumn{2}{c}{J1700-2610}
}
\startdata
&Continuum&$^{12}$CO $J=2-1$&$^{13}$CO $J=2-1$&C$^{18}$O $J=2-1$&SO $J_N=6_5-5_4$\\
\hline
Frequency (GHz)&225&230.5380000&220.3986842&219.5603541&219.9494420\\
Freq./vel. width&$\sim 2$ GHz&$0.32~\kms$&$0.17~\kms$&$0.17~\kms$&$0.17~\kms$\\
\hline
Beam (P.A.)&$0\farcs 29\times 0\farcs24\ (86\arcdeg)^{a}$&$0\farcs 35\times 0\farcs 25\ (76\arcdeg)^{a}$&$0\farcs 37\times 0\farcs 26\ (77\arcdeg)^{a}$&$0\farcs 38\times 0\farcs 27\ (78\arcdeg)^{a}$&$0\farcs 37\times 0\farcs 26\ (78\arcdeg)^{a}$\\
noise rms&$23~\uJB^{a}$&$1.4~\mJB^{a}$&$2.5~\mJB^{a}$&$1.9~\mJB^{a}$&$2.3~\mJB^{a}$\\
\hline
Beam (P.A.) &
\begin{tabular}{c}
$0\farcs 073\times 0\farcs048\ (77\arcdeg)^{c}$ \\ $0\farcs 043\times 0\farcs034\ (73\arcdeg)^{d}$ 
\end{tabular}
&$0\farcs 17\times 0\farcs 13\ (88\arcdeg)^{b}$&$0\farcs 17\times 0\farcs 13\ (87\arcdeg)^{b}$&$0\farcs 18\times 0\farcs 14\ (87\arcdeg)^{b}$&$0\farcs 18\times 0\farcs 14\ (87\arcdeg)^{b}$\\
noise rms&
\begin{tabular}{c}
$22~\uJB ^{c}$ \\ $100~\uJB ^{d}$
\end{tabular}
&$1.8~\mJB^{b}$&$3.0~\mJB^{b}$&$2.1~\mJB^{b}$&$2.5~\mJB^{b}$
\enddata
\tablecomments{The beam and noise rms values are shown for different robust parameters with superscripts of (a) Robust$=2$, (b) 0.5, (c) 0, and (d) $-2$.}
\end{deluxetable}

\section{Results} \label{sec:res}

\subsection{1.3 mm continuum} \label{sec:cont}
\begin{figure}[htbp]
\gridline{
\fig{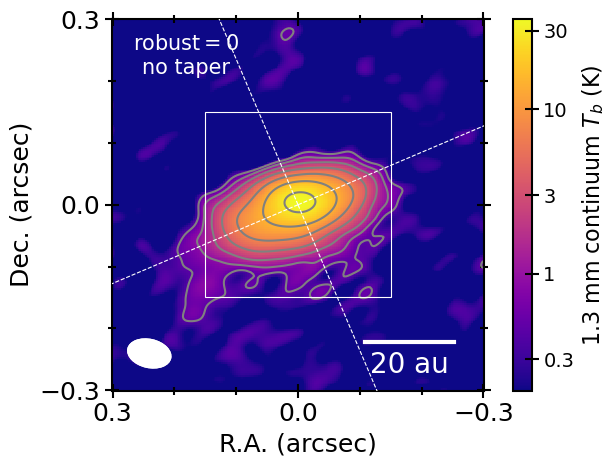}{0.33\textwidth}{(a)}
\fig{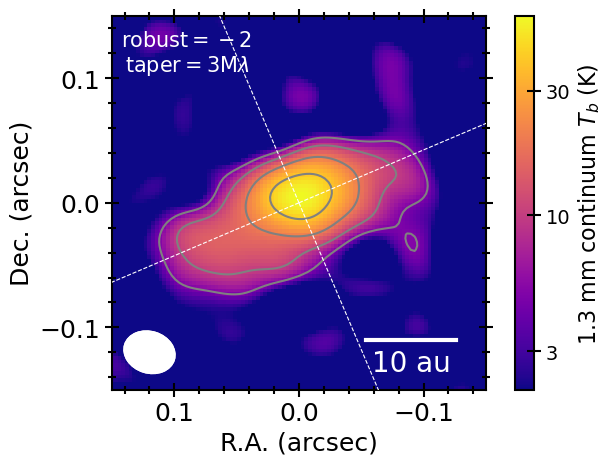}{0.33\textwidth}{(b)}
\fig{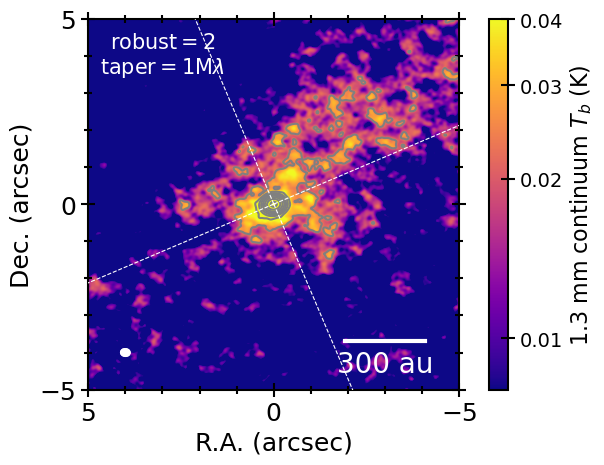}{0.33\textwidth}{(c)}
}
\caption{Images of the 1.3 mm continuum emission produced with (a) the robust parameter of 0.0 and no taper, (b) the robust parameter of $-2.0$ and a 3~M$\lambda$ ($\sim 0\farcs 06$) taper, (c) the robust parameter of 2.0 and a 1~M$\lambda$ ($\sim 0\farcs 18$) taper. The brightness temperature $T_b$ is calculated from the Rayleigh-Jeans approximation. The contour levels are $3,6,12,24,48,96,192\sigma$, where $1\sigma$ is (a) 0.022, (b) 0.100, (c) $0.023~\mJB$ (Table \ref{tab:obs}). The filled ellipses at the bottom-left corners denote the synthesized beams: (a) $0\farcs 073 \times 0\farcs 048$ ($77\arcdeg$), (b) $0\farcs 043 \times 0\farcs 034$ ($73\arcdeg$), and (c) $0\farcs 29 \times 0\farcs 24$ ($86\arcdeg$). The diagonal lines denote the major and minor axes of the emission in the image in panel (a): P.A.$=113\arcdeg$ and $23\arcdeg$. The square in panel (a) shows the plotted range of panel (b).
\label{fig:cont}}
\end{figure}

Figure \ref{fig:cont} shows continuum images with different robust and tapering parameters.
Figure \ref{fig:cont}(a) shows that the continuum emission with the robust parameter of 0 traces a disk-like structure on a $\sim 40$~au scale. This emission appears more extended to the southeast than to the northwest from the emission peak. This asymmetry is investigated in detail in Section \ref{sec:contsym}. Two dimensional (2D) Gaussian fitting to this robust$=0$ continuum image provides the central position of $(\alpha_{\rm ICRS},\ \delta_{\rm ICRS})=(16^{\rm h}28^{\rm m}21\fs 615,\ -24\arcdeg 36\arcmin 24\farcs 33$) with an uncertainty of (0.6 mas, 0.3 mas). The fitting, including the error bar calculation, used the CASA task {\it imfit}. The fitting result also provides the deconvolved size of $107\pm 2$ mas $\times$ $40\pm 2$ mas with a major axis in the direction of P.A. of $113\arcdeg \pm 1\arcdeg$. We adopt this direction as the major axis of the protostellar system, whici is perpendicular to the outflow direction, P.A.$=200\arcdeg -205\arcdeg$ \citep{yen17}. The aspect ratio corresponds to an inclination angle of ${\rm arccos}(40/107)\sim 68\arcdeg$, assuming a geometrically thin disk.

The peak intensity and flux density derived from the Gaussian fitting are $5.10\pm 0.02~\mJB$ and $11.5\pm0.2$~mJy, respectively. The flux densities integrated over $0\farcs 4 \times 0\farcs 4$ (the compact emission) and $10\arcsec \times 10\arcsec$ (the maximum recoverable scale of our ALMA observations) regions are $12.4\pm 0.3$ and $25\pm 2$~mJy, respectively. The flux density $F_{\nu}$ can be converted to a lower limit of gas mass, assuming that the emission is optically thin: $M_{\rm gas}=d^{2}F_{\nu}/\kappa_{\nu}B_{\nu}(T)$, where $d$, $\kappa _{\nu}$, and $B_{\nu}$ are the target distance, dust mass opacity, and the Planck function with an average temperature $T$, respectively. For this conversion, two different temperatures are adopted: {$T=20$ and 27~K.} $T=20$~K is derived from the relation between 850-$\micron$  continuum fluxes and model gas masses for 44 young stellar objects \citep{an.wi05}. $T=27$~K is calculated from the empirical relation, $T = 43~{\rm K}~(L_{\rm bol}/1 ~\Lsun)^{0.25}$, from \cite{tobi20}.
With these temperatures, $d=139$~pc, $\kappa_{\nu}=2.3~{\rm cm}^{2}~{\rm g}^{-1}$ \citep{beck90}, and a gas-to-dust mass ratio of 100, $F_{\nu}\simeq 12$~mJy corresponds to a gas mass of $M_{\rm gas}\simeq (1.4-2.1)\times 10^{-3}~\Msun$.

Figure \ref{fig:cont}(b) adopts the robust parameter of $-2$ and the 3-M$\lambda$ taper to focus on the central region with a reasonable S/N ratio. The plotted spatial range is half of Figure \ref{fig:cont}(a). This image shows an extension or a secondary peak to the southeast, separated from the central peak by $\sim 0\farcs 1$ roughly along the major axis. This is consistent with the extension seen in the image with robust$=0$ (Figure \ref{fig:cont}a).  

Figure \ref{fig:cont}(c), adopting the robust parameter of 2 and the 1-M$\lambda$ taper, shows the extended emission over $\sim 600$~au from the continuum emission peak to the northwest, as well as the compact ($\sim 100$~au) emission. This large-scale structure is also reported in \citet{yen17}.

\subsection{$^{12}$CO $J=2-1$} \label{sec:12co}

\begin{figure}[htbp]
\gridline{
\fig{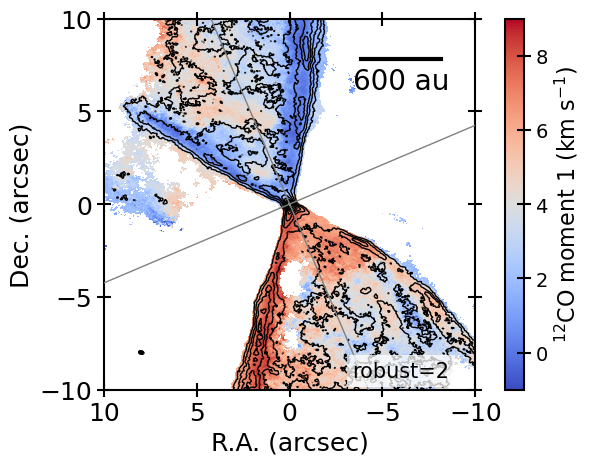}{0.49\textwidth}{(a)}
\fig{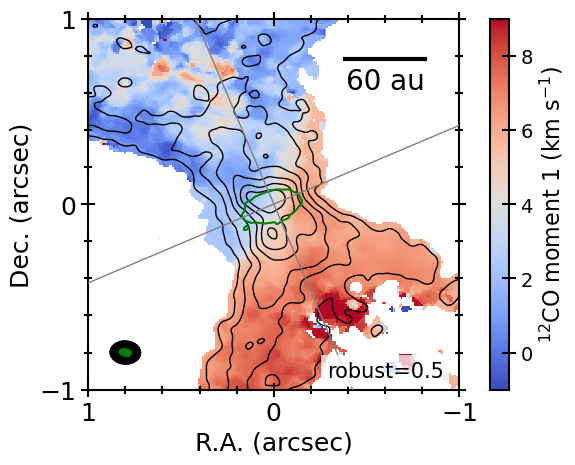}{0.49\textwidth}{(b)}
}
\caption{Moment 0 (integrated intensity) and 1 (mean velocity) maps in the $^{12}$CO $J=2-1$ line. The contour maps show the moment 0 map, while the color images show the moment 1 map. (a) Moment maps with the robust parameter of 2.0, integrated from $V_{\rm LSR}=-6.0$ to $14.0~\kms$. (b) Zoom-in view of the moment maps with the robust parameter of 0.5, integrated from $V_{\rm LSR}=-6.0$ to $14.0~\kms$. The contour levels are in (a) $12\sigma$ steps from $12\sigma$ with $1\sigma=3.2~\mJB~\kms$ and (b) $6\sigma$ steps from $6\sigma$ with $1\sigma=3.5~\mJB~\kms$. The filled ellipse at the bottom-left corner shows the synthesized beam: (a)  $0\farcs 35 \times 0\farcs 25$ ($76\arcdeg$) and (b)  $0\farcs 17 \times 0\farcs 13$ ($88\arcdeg$). The diagonal lines are the major and minor axes of the continuum emission, P.A.$=113\arcdeg$ and $23\arcdeg$. The green contour in panel (b) shows the 5$\sigma$ level of the continuum emission (Figure \ref{fig:cont}a).
\label{fig:12co}}
\end{figure}

Figure \ref{fig:12co}(a) shows the integrated intensity (moment 0) and the mean velocity (moment 1) maps in the $^{12}$CO emission produced with the robust parameter of 2. Positive and negative intensities are integrated for the moment 0 map, while intensities above the $3\sigma$ level are integrated for the moment 1 map. The noise level of each moment 0 map is calculated by the noise propagation from the noise level of the data cube (Table \ref{tab:obs}) used to make the moment 0 map: $\sigma _{\rm mom0}=\sigma _{\rm cube}dv\sqrt{N_{\rm ch}}$, where $\sigma _{\rm mom0}$, $\sigma _{\rm cube}$, $dv$, and $N _{\rm ch}$ are the noise level of the moment 0 map, the noise level of the cube data, the velocity width per channel (Table \ref{tab:obs}), and the number of integrated channels, respectively. The moment 0 and 1 maps of the other lines in this paper are also made in the same method. The $^{12}$CO emission traces a clear bipolar outflow whose axis is perpendicular to the major axis of the disk-like structure. An additional eastern component results from a large-scale structure around the systemic velocity, which is velocity resolved out by the interferometric observation. While the emission in the northern outflow lobe is overall blueshifted, redshifted emission can also be found on the northern side near the outflow axis. Similarly, the southern lobe is mostly redshifted but partly blueshifted. This is expected if a given outflow lobe crosses the plane of the sky that contains the central source, with the outflow axis close to the plane of sky \citep{cabr89}. A part of the lobe in the front of the plane of the sky appears blue-shifted, while the other part of the lobe behind the plane of the sky appears red-shifted. Using a radially-expanding parabolic outflow model, \citet{yen17} estimated the outflow shape and inclination by fitting a moment 0 map and a PV diagram along the outflow axis of the $^{12}$CO emission observed at an angular resolution of $\sim 1\arcsec$. Their best model provides an inclination angle of $60\arcdeg - 65\arcdeg$ ($0\arcdeg$ means pole-on) and an opening angle of $60\arcdeg -70\arcdeg$, calculated from the parabolic shape at the distance from the protostar along the outflow axis of $z\sim 5\arcsec - 10\arcsec$. This inclination is consistent with that of the disk-like structure detected in the dust continuum emission (Section \ref{sec:contsym}). Thus, their model is consistent with our observations which reveal blue- and red-shifted outflow components to the northeast and southwest, respectively.

Figure \ref{fig:12co}(b) shows a zoom-in view of the moment 0 and 1 maps with a different robust parameter of 0.5. The green contour shows the 5$\sigma$ level of the continuum emission (Figure \ref{fig:cont}a) for comparison. The moment 0 map shows local peaks within $\sim 0\farcs 4$ from the protostellar position, which is reported in \citet{hsie19} as a quadruple peak. In addition to the velocity gradient along the outflow axis on a $1\arcsec$ scale, this figure also shows a velocity gradient perpendicular to the outflow axis near the midplane on a $0\farcs 3$ scale, i.e., transition from a blueshifted part at P.A.$=113\arcdeg$ to a redshifted part at P.A.$=-67\arcdeg$. This velocity gradient is expected for rotating gas and analyzied in more detail in Section \ref{sec:kep}.

\begin{figure}[htbp]
\fig{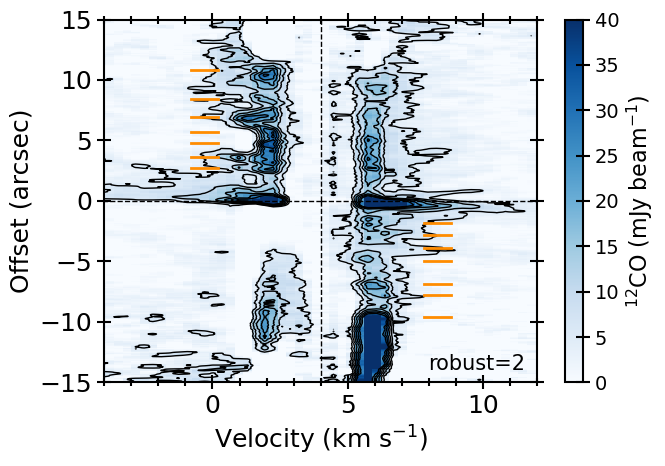}{0.7\textwidth}{}
\caption{Position-velocity diagram in the $^{12}$CO $J=2-1$ line along the outflow axis (P.A.$=23\arcdeg$) with a width of $0\farcs 6$. The positive offset corresponds to the northern side. The contour levels are in $6\sigma$ steps from $6\sigma$, where $1\sigma$ is $0.85~\mJB$. The dashed lines denote the systemic velocity ($V_{\rm LSR}=4~\kms$) and the protostellar position.
The orange short lines denote the position of local peaks or extensions visually identified.
\label{fig:pvco}}
\end{figure}

Figure \ref{fig:pvco} shows a position-velocity (PV) diagram of the $^{12}$CO emission along the outflow axis from the robust parameter 2 images and a width of $0\farcs 6$ ($\sim 2\times $ beam size). The width allows us to increase the S/N ratio but still focus on the velocity structure on the outflow axis. Because of this width, the noise level of this PV diagram is better than the one in Table \ref{tab:obs}. This figure also shows that both northern and southern lobes have both blue- and red-shifted parts. The outflow velocity is $|V-V_{\rm sys}|\sim 1.5-2.5~\kms$ at any position except for the central $1\arcsec - 2\arcsec$ region. The central component appears not to be a part of the outflow because of its $\sim 3$ times larger line widths. The outer ($>2\arcsec$) emission shows several ($\sim 5-9$) local peaks or extensions within $10\arcsec$ ($\sim 1400$~au) in each lobe, based on visual inspection (short lines in Figure \ref{fig:pvco}). This may be suggestive of episodic mass ejection \citep{lee20}. The interval angular scale, $10\arcsec/7=1.4\arcsec$, is significantly larger than the beam size of $0\farcs 35$. With the inclination angle of $65\arcdeg$ \citep{yen17}, the deprojected length $l$ and the deprojected velocity $v$ of the outflow can be calculated as $l=l'/\cos i$ and $v=v'/\sin i$, respectively, where $l'$ and $v'$ are the projected length and velocity. The interval $t$ of the mass ejection is a time required for material to move from a peak to the next peak. This can be calculated with the number of peaks $n$ within the length: $t=l'/n/v'$. Then, the values measured above provide
a typical interval of $1400~{\rm au}/\cos 65\arcdeg / (5\ {\rm to}\ 9) / (1.5\ {\rm to}\ 2.5~\kms / \sin 65\arcdeg)=600\ {\rm to}\ 2000$~year.
The episodic mass ejection is thought to be caused by episodic accretion bursts. The interval of accretion bursts is investigated in statistical studies using FUor objects. \citet{park21} reported accretion-burst intervals of $500-2300$~yr in a statistical study of FUor outbursts based on the fraction of outbursting sample in a set of NEOWISE observations. The fraction of variable sources is similar between VeLLOs like I16253 and more luminous objects in their sample. The uncertainty comes from the number of outbursts ranging from 2 to 9 and the total number of sample protostars ranging from 735 to 1059. This accretion-burst interval is consistent with the mass-ejection interval of I16253, implying that accretion bursts likely occurred in I16253.

This PV diagram shows a strong emission component at offsets of $<-10\arcsec$. This component is not confined around the outflow axis but extended over the redshifted lobe. This may imply that there could be more material on the southern side than on the northern side, although our observation did not target such a large spatial scale.

\subsection{C$^{18}$O and $^{13}$CO $J=2-1$}

\begin{figure}[htbp]
\gridline{
\fig{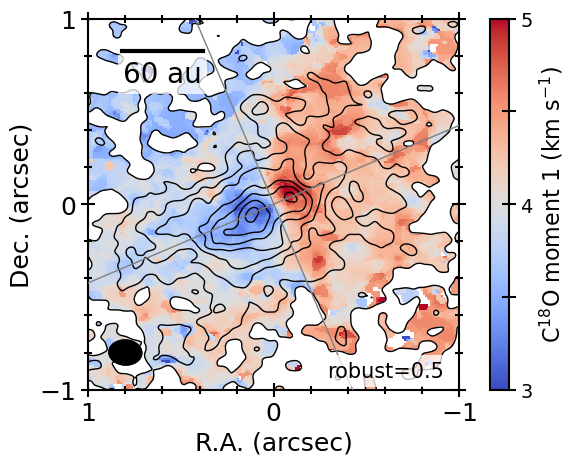}{0.4\textwidth}{(a)}
\fig{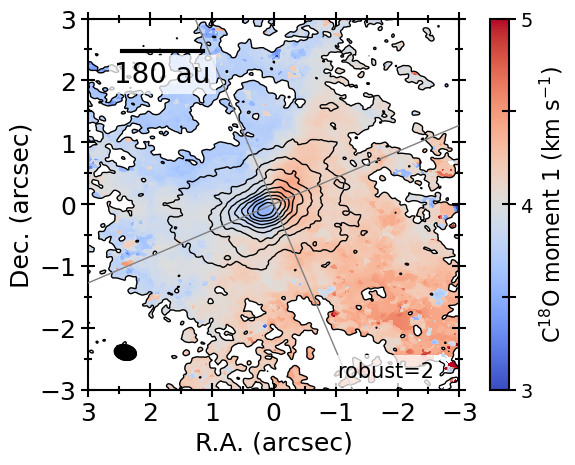}{0.4\textwidth}{(b)}
}
\gridline{
\fig{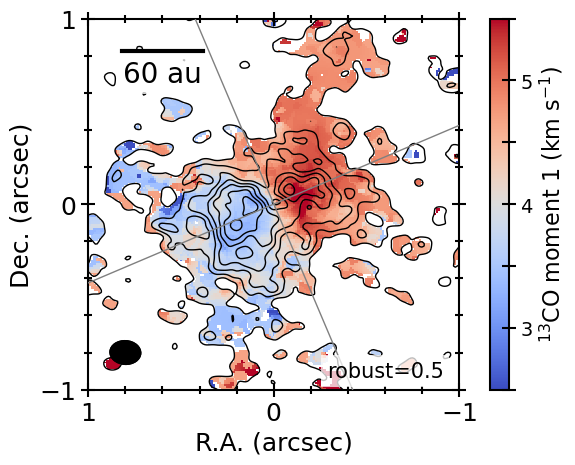}{0.4\textwidth}{(c)}
\fig{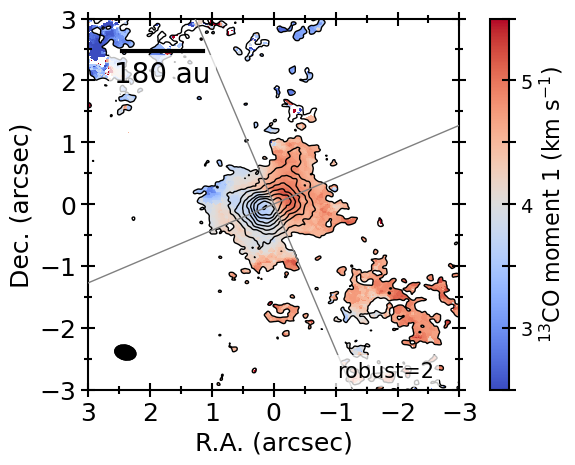}{0.4\textwidth}{(d)}
}
\caption{Moment 0 and 1 maps in the C$^{18}$O $J=2-1$ line emission and the $^{13}$CO $J=2-1$ line emission. The contour maps show the moment 0 map, while the color images show the moment 1 map. The contour levels are in $3\sigma$ steps in the robust$=0.5$ images and $6\sigma$ steps in the robust$=2$ images starting from $3\sigma$. The C$^{18}$O emission is integrated from 0.8 to $7.1~\kms$. These moment 0 maps have $1\sigma$ of 2.2 (robust$=0.5$) and $2.0~\mJB~\kms$ (robust$=2$). The $^{13}$CO emission is integrated from 0.7 to $7.3~\kms$. These moment 0 maps have $1\sigma$ of 2.8 (robust$=0.5$) and $2.7~\mJB~\kms$ (robust$=2$). The beam sizes are (a) $0\farcs 18 \times 0\farcs 14$ ($87\arcdeg$), (b) $0\farcs 38 \times 0\farcs 27$ ($78\arcdeg$) (c) $0\farcs 17 \times 0\farcs 13$ ($87\arcdeg$), and (d) $0\farcs 37 \times 0\farcs 26$ ($77\arcdeg$).
The diagonal lines are the major and minor axes of the continuum emission, P.A.$=113\arcdeg$ and $23\arcdeg$.
\label{fig:c18o13co}}
\end{figure}

Figure \ref{fig:c18o13co} shows the moment 0 and 1 maps in the C$^{18}$O and $^{13}$CO lines in robust$=0.5$ and 2. The two lines show an overall structure extended along the major axis (P.A.$=113\arcdeg$) and a velocity gradient primarily along the same direction. The integrated intensity maps with robust$=0.5$ show double peaks with a separation of $\sim 0\farcs 2-0\farcs 3$, and the integrated intensity is stronger at the eastern, blueshifted peaks than at the western, redshifted peaks, by $3\sigma$ in both lines. The integrated intensity maps with robust$=2$ show more clearly that the eastern emission is stronger than the western emission.
The detected emission size is larger in the C$^{18}$O line than in the $^{13}$CO line with both robust parameters. This is because the $^{13}$CO emission tends to be optically thicker and more extended than the C$^{18}$O emission, and thus the $^{13}$CO emission is resolved out more severely around the systemic velocity.

\subsection{SO $J_N=6_5-5_4$}

\begin{figure}[htbp]
\gridline{
\fig{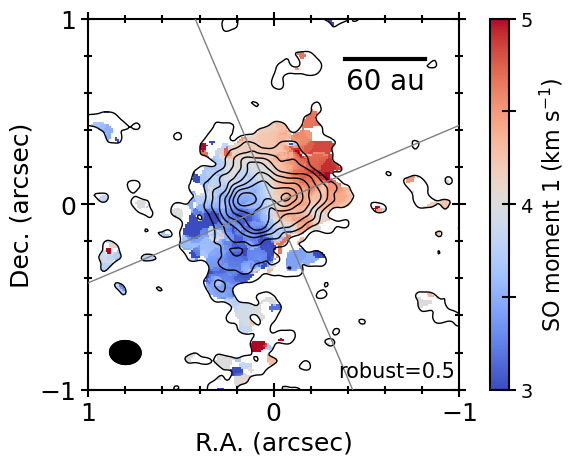}{0.4\textwidth}{(a)}
\fig{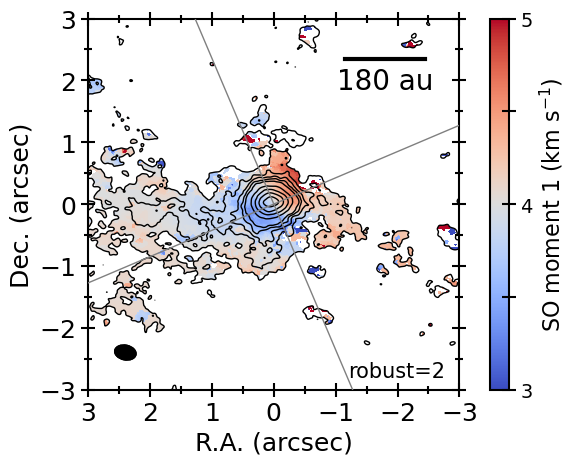}{0.4\textwidth}{(b)}
}
\caption{Moment 0 and 1 maps in the SO $J_N=6_5-5_4$ line emission. The contour maps show the moment 0 map, while the color images show the moment 1 map. (a) The robust parameter is 0.5. The emission is integrated from 1.7 to $6.3~\kms$. The contour levels are in $3\sigma$ steps from $3\sigma$ with $1\sigma=2.3~\mJB~\kms$. The beam size is $0\farcs 18 \times 0\farcs 14$ ($87\arcdeg$). (b) The robust parameter is 2. The emission is integrated from 1.7 to $6.3~\kms$. The contour levels are in $3\sigma$ steps from $3\sigma$ until $15\sigma$ and then in $10\sigma$ steps with $1\sigma=2.1~\mJB~\kms$. The beam size is $0\farcs 37 \times 0\farcs 26$ ($78\arcdeg$).
The diagonal lines are the major and minor axes of the continuum emission, P.A.$=113\arcdeg$ and $23\arcdeg$.
\label{fig:so}}
\end{figure}

Figure \ref{fig:so}(a) shows the integrated intensity and mean velocity maps in the SO line produced with robust$=0.5$ to focus on the central compact emission. The SO emission shows a double-peaked structure with a separation of $\sim 0\farcs 2-0\farcs 3$ and the velocity gradient along the major axis, like the C$^{18}$O and $^{13}$CO lines. On the other hand, the mean velocity of the SO emission is larger outside the peaks than inside the peaks, unlike the C$^{18}$O and $^{13}$CO lines. 

Figure \ref{fig:so}(b) shows the maps produced with robust$=2$ to focus on extended emission. The central double peaks are not spatially resolved in this map. The velocity structure in the central $1\arcsec$ region is overall the same as that in Figure \ref{fig:so}(a): eastern blueshifted emission and western redshifted emission. In addition, this figure shows an extended structure to the east from the central protostar. This structure is extended ($\sim 3\arcsec$) beyond the C$^{18}$O and $^{13}$CO emission ($\sim 1\arcsec$) in Figure \ref{fig:c18o13co}. An inner part ($\lesssim 1\arcsec$ from the protostellar position) of this extended structure shows velocities blueshifted from the systemic velocity, $V_{\rm LSR}=4~\kms$, which is consistent with the C$^{18}$O and $^{13}$CO velocities. In contrast, the outer part ($\gtrsim 1\arcsec$) shows slightly redshifted velocities ($V_{\rm LSR}-V_{\rm sys}\sim 0.3~\kms$). This redshifted velocity on the eastern side is different from the velocity gradient seen in the C$^{18}$O and $^{13}$CO lines and also different from the velocity in the eastern cavity wall of the $^{12}$CO outflow (Figure \ref{fig:12co}a). This SO component is discussed in more detail in Section \ref{sec:streamer}.

\section{Analysis} \label{sec:ana}

\subsection{Non-axisymmetry of the continuum image} \label{sec:contsym}

\begin{figure}[htbp]
\fig{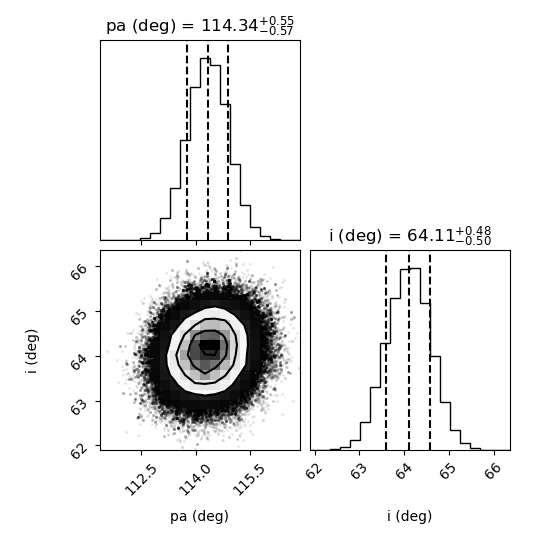}{0.5\textwidth}{}
\caption{Corner plot of the MCMC fitting to the continuum image. Two angles, $i$ and $pa$, are plotted among the 14 free parameters. The dashed lines show the 16, 50, and 84 percentiles.
\label{fig:corner}}
\end{figure}

The 1.3 mm continuum emission with robust$=0$ (Figure \ref{fig:cont}a) is more extended in the southeastern side ($\sim 30$~au) than in the northwestern side ($\sim 20$~au). We verify that the extended components are real by comparing images made with various robust parameters. To extract the extended component, an axisymmetric model of the dust continuum emission is constructed in this subsection. The model is made from a radial profile of intensities in the unit of Jy~pixel$^{-1}$, before beam convolution, set with free parameters of $I(r=0), I(dr), I(2dr), ..., I(11dr)$. The grid separation $dr$ is half of the beam minor-axis. The largest radius $11dr=0\farcs 26$ covers the entire continuum emission with robust$=0$. Then, the circular 2D intensity distribution is projected by the inclination factor of $\cos i$ in the direction of the minor axis and rotated by the position angle $pa$; these two angles are also free parameters. This elliptical 2D intensity in the unit of Jy~pixel$^{-1}$ is convolved with the observational beam to make the model image in the unit of $\JB$. Hence, this model does not assume any shape, such as a Gaussian function but only the axisymmetry. This modeling method follows the one in \citet{aso21}. The central position is not a free parameter but fixed at the observed peak position, $(\alpha _{\rm ICRS},\ \delta _{\rm ICRS})=(16^{\rm h}28^{\rm m}21\fs 6153,\ -24\arcdeg 36\arcmin 23\farcs 325)$, which coincides with the Gaussian center derived in Section \ref{sec:cont}. The best-fit parameters are obtained by minimizing $\chi ^2$ between the observed and model images, divided by the number of pixels in the beam, through the Markov chain Monte Carlo (MCMC) methods with the public Python package \texttt{ptemcee}. The numbers of free parameters, steps, {\rm and} walkers per parameter are 14, 8000, and 16, respectively. The first half (4000) steps were removed for the burn-in. The derived best-fit angles are $i=64.1\arcdeg \pm 0.5\arcdeg$ and $pa=114.3\arcdeg \pm 0.6\arcdeg$, and Figure \ref{fig:corner} shows the posterior distribution of these two angles produced by the MCMC fitting. The uncertainties of these angles are derived as the 16 and 84 percentiles.

\begin{figure}[htbp]
\gridline{
\fig{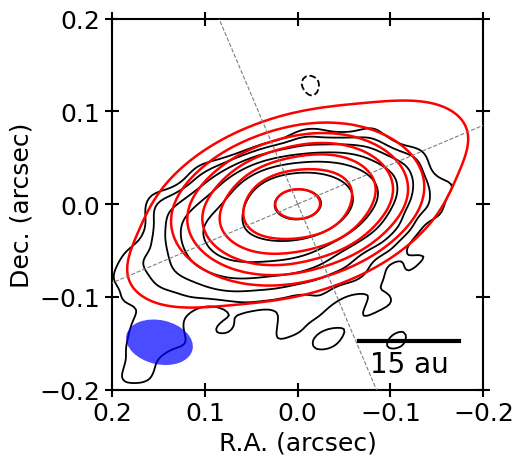}{0.49\textwidth}{(a)}
\fig{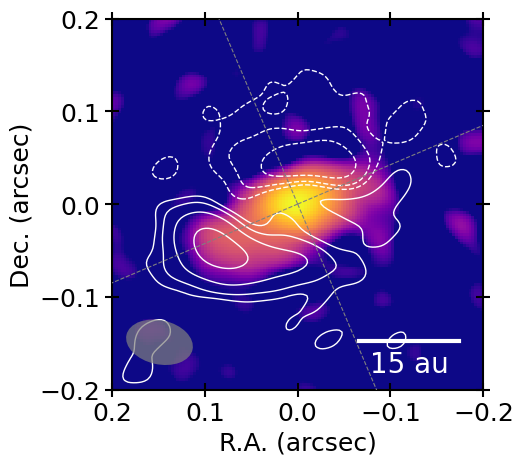}{0.49\textwidth}{(b)}
}
\caption{(a) Axisymmetric model (red) fitted to the observed 1.3 mm continuum emission (black) with the robust parameter of 0.0. The contour levels are the same as Figure \ref{fig:cont}(b): 3,6,12,24,48,96,192$\sigma$. (b) Residual (contour map) between the model and observation in panel (a), overlaid on the robust$=-2$ continuum image (color map; Figure \ref{fig:cont}c). The contour levels are the same: $3,6,12,24\sigma$.
\label{fig:contmodel}}
\end{figure}

Figure \ref{fig:contmodel}(a) shows the comparison between the best-fit axisymmetric model and the observed image. The overall structure in the observed image is reproduced, but the model emission is weaker than the observed emission in the southeast, while it is stronger in the northwest, as expected. The contour map in Figure \ref{fig:contmodel}(b) shows the residual image after subtracting the model from the observed image with robust$=0$, while the color map shows the observed continuum image with robust$=-2$ (same as Figure \ref{fig:cont}c). The strong residual is located in the southeast. This residual overlaps with the extension in the robust$=-2$ image, supporting that this excess from the symmetric component is not due to the specific robust parameter. The direction of the residual and extension coincides with the direction of the SO extended structure and the stronger peak on the eastern side in the C$^{18}$O and $^{13}$CO moment 0 maps.

\subsection{Keplerian rotation in the $^{12}$CO emission} \label{sec:kep}
\begin{figure}[htbp]
\gridline{
\fig{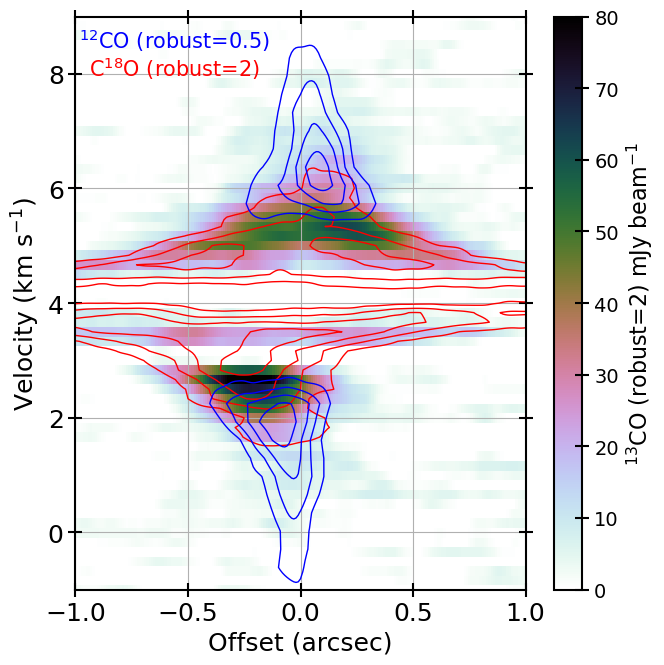}{0.49\textwidth}{(a)}
\fig{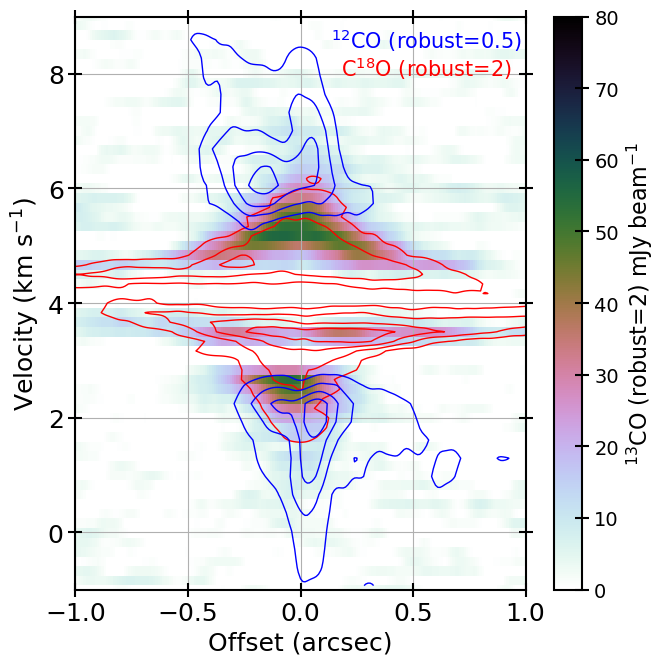}{0.49\textwidth}{(b)}
}
\caption{Position-velocity diagrams in the $^{12}$CO (blue; robust$=0.5$), C$^{18}$O (red; robust$=2.0$), and $^{13}$CO (color; robust$=2.0$) $J=2-1$ lines along the (a) major and (b) minor axes. The positive offset corresponds to the western and northern sides in panels (a) and (b), respectively. The contour levels are in $6\sigma$ steps from $6\sigma$, where $1\sigma$ is $1.4~\mJB$ for $^{12}$CO, $2.1~\mJB$ for C$^{18}$O, and $3.0~\mJB$ for $^{13}$CO.
The beam sizes in these diagrams are $\sim 0\farcs 15$ for $^{12}$CO, $\sim 0\farcs 33$ for C$^{18}$O, and $\sim 0\farcs 32$ for $^{13}$CO (Table \ref{tab:obs}).
\label{fig:pv13co}}
\end{figure}

Previous observational studies have not yet identified any part of gas motion in I16253 having a radial profile of the Keplerian rotation but assumed that the observed gas follows Keplerian rotation. The Keplerian disk is crucial to estimate the central stellar mass in the protostellar phase and verify whether I16253 is a proto-BD or a very low-mass protostar. To tackle this problem, we analyze the position-velocity (PV) diagrams along the major axis. Figure \ref{fig:pv13co}(a) shows the PV diagrams in the $^{12}$CO (blue contours), C$^{18}$O (red contours), and $^{13}$CO (color) lines. The C$^{18}$O and $^{13}$CO lines show a typical shape of an infalling plus rotating motion: a distorted diamond shape with emission in all the four quadrants, as reported in \citet{hsie19}. In contrast, the $^{12}$CO emission is concentrated on the first (upper right) and third (lower left) quadrants (the western redshifted emission and the eastern blueshifted emission), showing a clear velocity gradient, in $\pm (0\farcs 1-0\farcs2)$ at $|V-V_{\rm sys}|\gtrsim 2~\kms$. Figure \ref{fig:pv13co}(b) shows the PV diagrams in the three lines along the minor axis. The C$^{18}$O and $^{13}$CO emission can be seen in all four quadrants in the minor-axis PV diagram with a diamond shape, which is typical for the infall motion \citep[e.g.,][]{ruiz22}. The $^{12}$CO emission is mainly concentrated around the center, with additional components in the second and fourth quadrants (the northern blueshifted emission and the southern redshifted emission). While the additional components show a velocity gradient due to the outflow motion, the main component shows no velocity gradient along the minor axis. These PV diagrams suggest that the $^{12}$CO line traces a purely rotating, i.e., Keplerian disk, while the C$^{18}$O and $^{13}$CO lines trace the infalling rotating envelope previously identified. This difference among individual molecular lines can be understood by the different strengths of emission and the different missing fluxes in the interferometric observations. The $^{12}$CO emission is strong enough to be detected even in the expected compact disk ($r\sim 0\farcs 1-0\farcs 2$ in the PV diagram as well as in the continuum image). The $^{13}$CO and C$^{18}$O emission is not detected sufficiently in this compact region at the high velocities where the $^{12}$CO emission is detected. The nondetection in the $^{13}$CO and C$^{18}$O lines suggests the $^{12}$CO emission is not optically thick at the high velocities. In contrast to the high velocities, the $^{12}$CO emission is more extended and strongly resolved out at the low velocities where the $^{13}$CO and C$^{18}$O emission traces the envelope.

In order to verify whether the $^{12}$CO line traces the Keplerian rotation, we first find the emission ridge \citep[e.g.,][]{yen13,aso15,sai20} and edge \citep[e.g.,][]{seif16,alve17} in the major-axis PV diagram and fit them with power-law functions. The ridge is a center of the emission along the positional axis at each velocity, while the edge is an outer boundary of the emission along the positional axis at each velocity. The analysis and fitting process was performed through the Python open package {\tt pvanalysis} in 
{\tt Spectral Line Analysis/Modeling} \citep[{\tt SLAM};][]{as.sa23}\footnote{\url{https://github.com/jinshisai/SLAM}},
and the detail is described in the overview paper of the eDisk project \citep{ohas23}. 
We thus mention here the settings specific to the case of I16253. The ridge point is defined as the intensity-weighted mean position $x_m$ at each velocity $v$: $x_m(v)=\int I(x, v)xdx/\int I(x, v)dx$. $x_m$ is calculated using the intensities above the $5\sigma$ level along the positional axis at each velocity in the range of $|V-V_{\rm sys}|>2~\kms$. This velocity range is selected because the emission is located in the first and third quadrants in this velocity range. Similarly, using the 1D intensity profile, the edge point is defined, at each velocity in the same velocity range, as the outer position where the intensity is at the $5\sigma$ level. In addition, the tool uses velocities higher than or equal to the velocity at which the derived edge/ridge radius is largest so that the relation between radius and velocity can be consistent with spin-up rotation. The ridge and edge tend to under- and overestimate, respectively, the radius at a given velocity \citep[][the ridge and edge here correspond to their ``centroid'' and ``first emission contour'', respectively]{mare20}, and thus we adopt the former and the latter as the lower and upper limits of radius in this paper.

\begin{figure}[htbp]
\gridline{
\fig{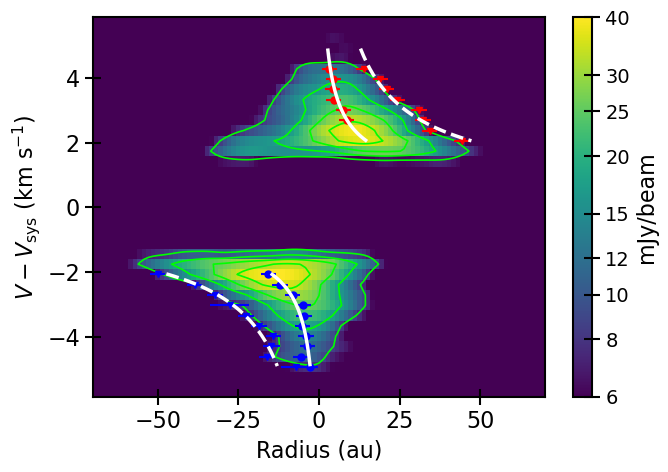}{0.49\textwidth}{(a)}
\fig{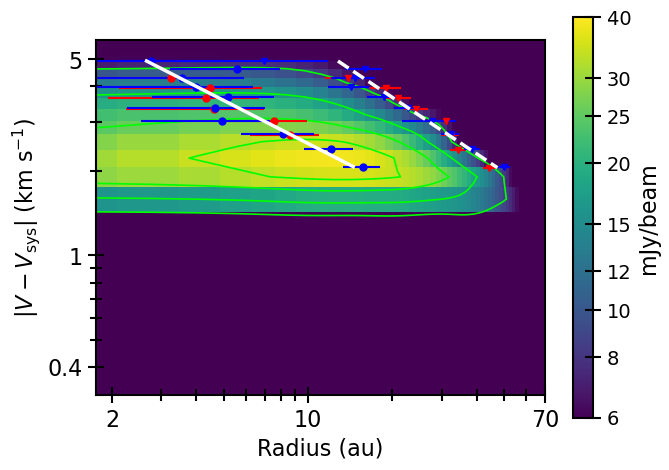}{0.49\textwidth}{(b)}
}
\caption{Edge and ridge points, estimated for each velocity channel, overlaid on the $^{12}$CO major-axis PV diagram (robust$=0.5$) in the (a) linear and (b) logarithmic scales. The positive radius corresponds to the western side. The velocity is the relative velocity to the systemic velocity of $V_{\rm LSR}=4~\kms$. The white curves/lines are the best-fit power-law functions (Section \ref{sec:kep}). The contour levels are in the $3\sigma$ steps from $3\sigma$, where $1\sigma$ is $1.4~\mJB$.
\label{fig:pvanaco}}
\end{figure}

Figure \ref{fig:pvanaco} shows the estimated ridge and edge points overlaid on the PV diagram in the linear and logarithmic scales. The logarithmic diagram is made by avaraging the emission in the first and third quadrants of the linear diagram. The edge and ridge radii are separately fitted with a power-law relation between either edge or ridge radius $R$ and the velocity $V$: 
\begin{eqnarray}
V = V_b \left(\frac{R}{R_b}\right)^{-p}+V_{\rm sys}, \label{eq:pow}\\
p = p_{\rm in}\ {\rm if}\ R<R_b\ {\rm else}\ p_{\rm in}+dp.
\end{eqnarray}
This fitting uses the MCMC method with the tool of {\it emcee}, where the numbers of walker per free parameter, burn-in steps, adopted steps are 16, 2000, and 1000, respectively. The error bar of each free parameter is defined as the 16 and 84 percentiles.  First, we adopt a single power-law function, where the free parameters are the power-law index $p$ and the radius $R_b$ at a fixed middle velocity $V_b$. $dp=0$ and $V_{\rm sys}=4.0~\kms$ are fixed. The fixed parameter $V_b$ is an arbitrary reference velocity where $R_b$ is derived from the fitting. The central stellar mass $M_*$ is calculated from the radius with the fixed inclination angle $i=65\arcdeg$ as $M_* =R(V) (V/\sin i) ^2 / G$, where $R(V)$ is calculated from Equation (\ref{eq:pow}) as a function of $V$. The calculated $M_*$ can thus depend on $V$ if $p\neq 0.5$. The uncertainty of $M_*$ is calculated through the error propagation. The inclination angle is adopted from an outflow model \citep[$60\arcdeg-65\arcdeg$;][]{yen17} and our continuum analysis ($64.3\arcdeg$ in Section \ref{sec:contsym}). Even with $i=60\arcdeg$, the calculated stellar mass is only 10\% higher. The best-fit parameters are $p=0.5$ and $M_*=0.09~\Msun$ with the ridge points, while they are $p=0.67$ and $M_* = 0.3-0.4~\Msun$ with the edge points. The range of edge $M_*$, $0.3-0.4~\Msun$ is because $p_{\rm in}\neq 0.5$. This range is wider than the statistical uncertainty. The fitting results are summarized in Table \ref{tab:power} including statistical uncertainties. The uncertainty of the distance adds a relative uncertainty of $\sim 7\%$. The best-fit functions are plotted in Figure \ref{fig:pvanaco}. Figure \ref{fig:pvanaco}(a) shows that the best-fit functions trace the ridge and edge of the observed PV diagram. Figure \ref{fig:pvanaco}(b) clearly shows that the ridge and edge indices are close to each other, as well as to that of Keplerian rotation $(p=0.5)$. The ridge and edge points were also fitted with a double power-law function. The best-fit parameters are summarized in Table \ref{tab:power}. The double power-law fitting yields similar power-law indices $p=0.5-0.6$ for the most part of the fitted velocity range, except for only the lowest velocity channel. These results indicate that the $^{12}$CO line traces the Keplerian disk around I16253.

\begin{deluxetable}{cccccc}
\tablecaption{Power-law fitting to the edge and ridge points in the major-axis $^{12}$CO PV diagram. \label{tab:power}}
\tablehead{
 & \multicolumn{5}{c}{Single power}
}
\startdata
& $R_b$ (au) & $V_b$ ($\kms$) & $p_{\rm in}$ & $dp$ & $M_* (\Msun)$\\
Edge & $24.7\pm 0.7$ & 3.18 (fixed) & $0.67\pm 0.04$ & 0 (fixed) & $0.34\pm0.01-0.43\pm 0.02$ \\
Ridge & $6.2\pm 0.7$ & 3.20 (fixed) & $0.5\pm 0.1$ & 0 (fixed) & $0.09\pm 0.02$\\
\hline
 & \multicolumn{5}{c}{Double power}\\
\hline
& $R_b$ (au) & $V_b$ ($\kms$) & $p_{\rm in}$ & $dp$ & $M_* (\Msun)$\\
Edge & $45\pm 10$ & $2.2\pm 0.5$ & $0.63\pm 0.07$ & $2\pm 4$ & $0.3\pm 0.2-0.4\pm 0.2$\\
Ridge & $13\pm 2$ & $2.3\pm 0.2$ & $0.5\pm 0.1$ & $4\pm 4$ & $0.09\pm 0.04$\\
\enddata
\tablecomments{The uncertainties of $R_b$ and $M_b$ in this table are before the uncertainty of the distance, $\sim 7\%$, is incorpolated.}
\end{deluxetable}

By identifying the Keplerian rotation, we have estimated the central stellar mass to be $M_*=0.09-0.34~\Msun$ directly (kinematically) for the first time. Even with the lower limit, I16253 already has sufficient mass to evolve into a low-mass star, beyond the brown-dwarf mass regime ($M_* > 0.08~\Msun$). \citet{mare20} demonstrated that the central stellar masses derived from the ridge and edge points under- and overestimate the actual stellar mass by $\sim 30\%$ and $\sim 100\%$ in the case where the beam size is a few tens percent larger than the observed disk size. The reason why the ridge underestimates the stellar mass is that a part of the gas inside the Keplerian radius has the same line-of-sight velocity as the gas at the Keplerian radius, and this inner emission shifts the ridge inward. This effect is also discussed quantitatively by \citet{aso15}. The reason why the edge overestimates the stellar mass is that the observational beam causes emission outside the outermost, i.e., Keplerian radius, and this outer emission shifts the edge outward. The condition in \citet{mare20} is applicable to our case (the disk radius of I16253 is discussed in Section \ref{sec:diskringenv}). Then, the central stellar mass is likely within $M_*=0.09\times 1.3$ to $0.34/2~\Msun=0.12$ to $0.17~\Msun$. This mass is consistent with a value derived from the $^{12}$CO $J=2-1$ emission at two velocity channels by assuming  Keplerian rotation \citep[$\sim 0.12~\Msun$;][]{hsie19}.

\subsection{Linear velocity gradient in the SO emission} \label{sec:anaso}
\begin{figure}[htbp]
\fig{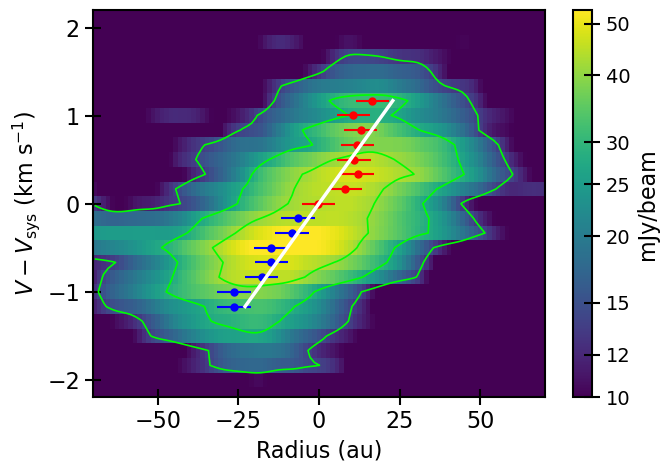}{0.7\textwidth}{}
\caption{Ridge points, found for each velocity, overlaid on the SO PV diagram (robust$=2$) along the major axis in the linear scale. The positive radius corresponds to the western side. The velocity is relative to the systemic velocity of $V_{\rm LSR}=4~\kms$. The white line is the best linear function (Section \ref{sec:anaso}). The contour levels are in the $3\sigma$ steps from $3\sigma$, where $1\sigma$ is $2.3~\mJB$.
\label{fig:pvanaso}}
\end{figure}

Figure \ref{fig:pvanaso} shows the PV diagram in the SO emission along the major axis. The SO emission is detected within $|V-V_{\rm sys}|\sim \pm 2~\kms$, which is similar to the velocity range where the $^{13}$CO and C$^{18}$O emission is present (see Figure \ref{fig:pvrgb} for the comparison of the SO and C$^{18}$O emission). The shape of the SO PV diagram is represented by a linear velocity gradient from the eastern blueshifted emission to the western redshifted emission in the velocity range of $|V-V_{\rm sys}|<1.3~\kms$, unlike the CO isotopologues. The SO emission outside this velocity range appears to overlap a component traced by the CO isotopologue emissions (see Figure \ref{fig:pvrgb} for the comparison), although it is difficult to discuss it given the limited number of the channels at the common velocity range. The linear velocity gradient can also be seen in the moment 1 map (Figure \ref{fig:so}a), where the high velocities are located at the outermost parts along the major axis. To evaluate the linear velocity gradient quantitatively, the ridge points are found at each velocity, as plotted in Figure \ref{fig:pvanaso}, and fitted with a linear function of radius $r$ passing $(r, v)=(0, 0)$, using {\it emcee} with the same condition as that in Section \ref{sec:kep}. The best-fit gradient is estimated to be $g=0.056\pm 0.005~\kms~{\rm au}^{-1}$, after the correction for the inclination angle of $i=65\arcdeg$. This value is slightly smaller than a previous result of $0.082\pm 0.004~\kms~{\rm au}^{-1}$ ($d=139$~pc) by \citet{yen17}. This difference is mainly due to the different angular resolutions, $\sim 0\farcs2$ in the present observations and $\sim 1\arcsec$ in \citet{yen17}. Their result may thus include SO emission at larger radii and slower velocities.
The SO emission is located within the velocities of $|V-V_{\rm sys}|<2~\kms$, which are lower than those showing the Keplerian rotation in the $^{12}$CO PV diagram (Figure \ref{fig:pvanaco}), implying that the SO emission traces a different part from the $^{12}$CO emission. The SO velocity structure and the velocity gradient are discussed in more detail in Section \ref{sec:diskringenv} along with the infalling and rotating motions traced in the C$^{18}$O and $^{13}$CO emission.

\section{Discussion} \label{sec:dis}

\subsection{Disk, Ring, and Envelope} \label{sec:diskringenv}
\begin{figure}[htbp]
\gridline{
\fig{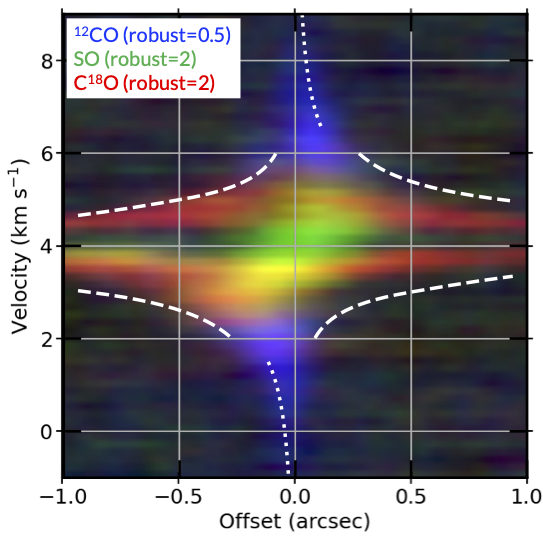}{0.49\textwidth}{(a)}
\fig{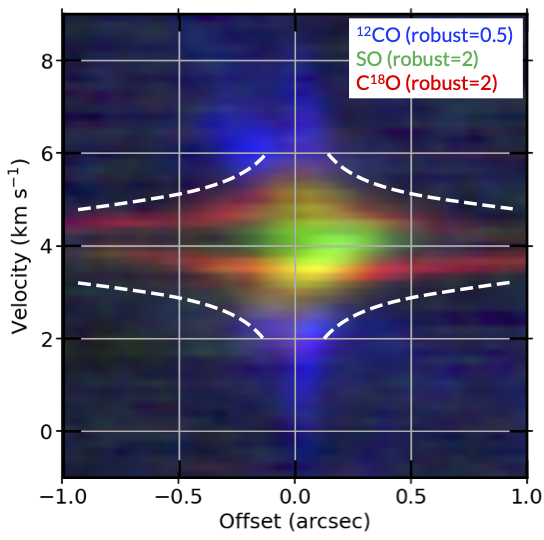}{0.49\textwidth}{(b)}
}
\caption{Comparison between the predicted maximum line-of-sight velocities (white dashed and dotted curves) and the observed PV diagrams (C$^{18}$O in red, SO in green, and $^{12}$CO in blue). (a) PV diagram along the major axis. The dashed curves represent the UCM envelope model with a central stellar mass of $M_*=0.14~\Msun$ (Section \ref{sec:kep}) and the specific angular momentum $j=45~\kms~{\rm au}$ \citep{hsie19}. The envelope maximum velocity is calculated only up to $|V-V_{\rm sys}|=2~\kms$. The dotted curves show the maximum line-of-sight velocities of the Keplerian disk with a disk radius of 16~au. (b) PV diagram along the minor axis. The dashed curves represent the same UCM envelope as for the major axis.
\label{fig:pvrgb}}
\end{figure}

We suggest a picture of the I16253 system based on our results and the previous results. \citet{hsie19} estimated the specific angular momentum of the envelope to be $j\sim 45~\kms {\rm au}$ ($d=139~pc$), by reproducing the PV diagrams along the major and minor axes in the C$^{18}$O and $^{12}$CO lines with an infalling and rotating envelope model. When the specific angular momentum and the central stellar mass are given, the velocity field of a protostellar envelope can be predicted by the UCM envelope model \citep{ulri76, ca.mo81}, where the velocity field consists of ballistic, parabolic flows from an outer boundary in the rigid-body rotation. Figure \ref{fig:pvrgb} compares the maximum line-of-sight velocity of the UCM envelope model, as well as that of the Keplerian disk model, at each position and the observed PV diagrams in the C$^{18}$O, SO, and $^{12}$CO lines along the major and minor axes. The model parameters are the specific angular momentum of $j\sim 45~\kms {\rm au}$ and the central stellar mass of $M_*\sim 0.14~\Msun$ (middle of $0.12-0.17~\Msun$ in Section \ref{sec:kep}). With the specific angular momentum and the central stellar mass, the centrifugal radius is calculated to be $R_c =j^2 / GM_{*}\sim 16$~au ($0\farcs 12$); $M_*=0.12-0.17~\Msun$ corresponds to $R_c=13-19$~au. $R_c=16$~au is used as the disk radius to draw the model curves in Figure \ref{fig:pvrgb}. The C$^{18}$O PV diagrams are consistent with the model maximum velocity both in the major and minor axes in the velocity range lower than the disk velocities, $|V-V_{\rm sys}|<2~\kms$. The obtained $R_c$ is close to the radius of the continuum emission (the red $6\sigma$ contour in Figure \ref{fig:contmodel}a), except for the southeastern extension, supporting that this is the disk radius. This is also consistent with a relation suggested by \citet{as.ma20} that the Gaussian deconvolved radius of the 1.3 mm continuum emission ($\sim 7.4$~au for I16253; Section \ref{sec:cont}) is $\sim 0.5$ as large as the disk radius for an evolutionary phase from $M_*\sim 0.1$ to $0.4~\Msun$; this relation is based on their synthetic observations of a magnetohydrodynamics simulation of protostellar evolution.

The SO emission shows a different shape from the envelope and from the disk in the major-axis PV diagram, whereas the detected velocity range is similar to that of the C$^{18}$O (and $^{13}$CO) emission as shown in Figure \ref{fig:pvrgb}. The linear velocity gradient in the major-axis PV diagram, along with the double peak structure in the moment 0 map at a higher angular resolution (Figure \ref{fig:so}a), suggests that the SO emission traces a ring close to edge-on due to accretion shock between the infalling envelope and the disk, as reported in other protostellar systems \citep[e.g.,][]{yen14, ohas14, saka16}. The linear velocity gradient in the SO PV diagram $0.056~\kms~(r / 1~{\rm au})$ intersects with the envelope rotation $45~\kms~(r / 1~{\rm au})^{-1}$ at $r=28$~au. This radius is close to the outermost radius of the SO ridge points (Figure \ref{fig:pvanaso}).
Recent theoretical studies predict that the accretion shock around the disk occurs at the radius of $\sim 1.5R_c$ \citep{shar22}. The inferred radius of the SO ring in I16253 is approximately consistent with this prediction. In conclusion, we suggest that I16253 has the Keplerian disk with $r\sim 16\pm 3$~au traced in the $^{12}$CO line, which is surrounded by the shock at $r\sim 28$~au traced in the SO line due to the mass accretion from the envelope traced in the C$^{18}$O and $^{13}$CO lines.

\subsection{Streamer in the SO emission} \label{sec:streamer}
\begin{figure}[htbp]
\fig{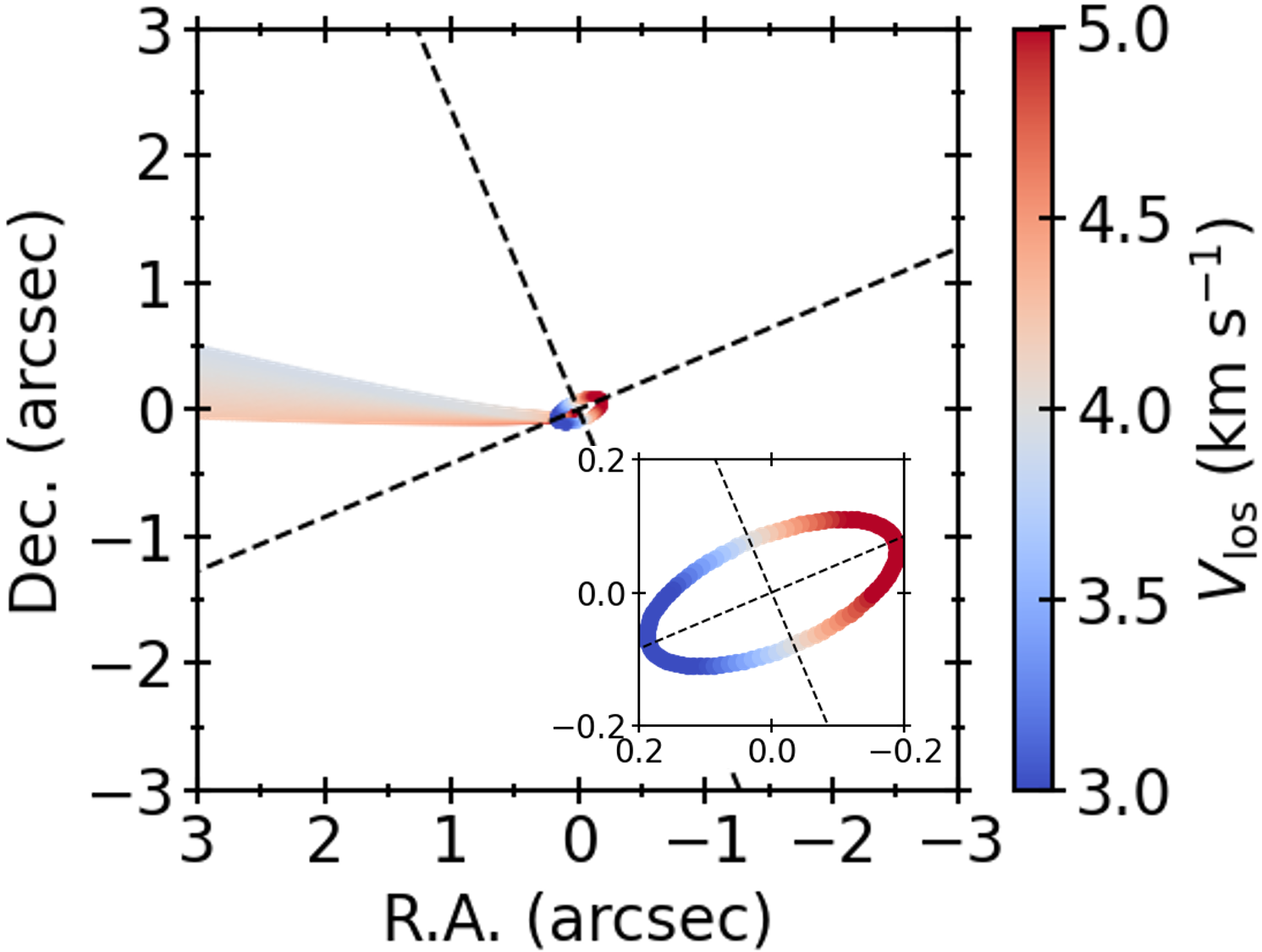}{0.4\textwidth}{}
\caption{Model map of the line-of-sight velocity of a streamer in the UCM envelope ($M_* = 0.14~\Msun$ and $j=45~\kms ~{\rm au}$ are the same as for Figure \ref{fig:pvrgb}) and a Keplerian ring at 28~au. The streamer comes from $30\arcdeg$ above the midplane (i.e., $\theta _0=60\arcdeg$) and $0\arcdeg - 20\arcdeg$ away from the left to the front seen from the observer around the rotational axis (i.e., $\phi _0=240\arcdeg - 270\arcdeg$).
\label{fig:streamer}}
\end{figure}

In addition to the shocked ring, the SO emission also shows an extended structure to the east as shown in Figure \ref{fig:so}(b). Its slightly redshifted velocity in the outer ($\gtrsim 1\arcsec$) region cannot be explained by the rotation, the infall motion on the midplane nor the outflow seen in our observations toward I16253, implying a different motion. A possible explanation is a streamer as reported in other protostellar systems on $\lesssim 1000$~au scales \citep[e.g.,][]{yen19, garu22, thie22} \citep[see also][]{kido23} as well as on larger scales \citep[e.g.,][]{pine20}. Hence, we constructed a streamer model by extracting ballistic, parabolic flows from the UCM envelope model used in Section \ref{sec:diskringenv}. The free parameters are the two directional angles $(\theta_0, \phi _0)$ to specify the initial polar and azimuthal angle of the streamer trajectory, respectively. The polar angle $\theta _0$ is $0\arcdeg$ at the disk northern pole and $90\arcdeg$ at the midplane. The azimuthal angle $\phi _0$ is $0\arcdeg$ in the direction of the observer on the midplane and $90\arcdeg$ on the right seen from the observer. The inclination and position angles of the midplane are fixed at $65\arcdeg$ and $113\arcdeg$, respectively. By visual inspection, we found that $(\theta _0, \phi _0)=(60\arcdeg, 240\arcdeg-270\arcdeg)$ can reasonably reproduce the observed SO distribution and velocity. This suggests that the SO gas is gravitationally bound if it came from these $(\theta _0, \phi _0)$ angles, since the UCM envelope model describes how material infalls toward the central gravitational source.
Figure \ref{fig:streamer} shows the distribution of the model streamer projected on the plane-of-sky and the line-of-sight velocity of the streamer model and the {\rm rotating} ring at a radius of 28~au. The ring rotates in the same velocity as the UCM envelope model. The model ring appears to reproduce the size and line-of-sight velocity of the observed compact component (Figures \ref{fig:so}a), while the model streamer appears to reproduce the extended structure to the east and its line-of-sight velocity (Figure \ref{fig:so}b). Meanwhile, the observed extended structure appears less confined than the model. This is probably because our model did not consider any beam blurring effect or radiative transfer effect. More sophisticated modeling including such effects will help to verify whether the anisotropy in the SO line is caused by the streamer.

Ballistic streamers are simulated in a theoretical study of the cloudlet capture in a protostellar system \citep{hana22}. The simulation shows that the streamers can be recognized as enhanced molecular line emission. Such an enhancement is seen in the eastern peak ($3\sigma$ higher than the western peak) of the C$^{18}$O and $^{13}$CO moment 0 maps in I16253 (Figure \ref{fig:c18o13co}), as well as in the SO extended emission. The eastern extension in the continuum emission (Figures \ref{fig:cont}b and \ref{fig:contmodel}b) could also be explained by the enhancement due to the streamer since it is located on the eastern side.

\begin{figure}[htbp]
\fig{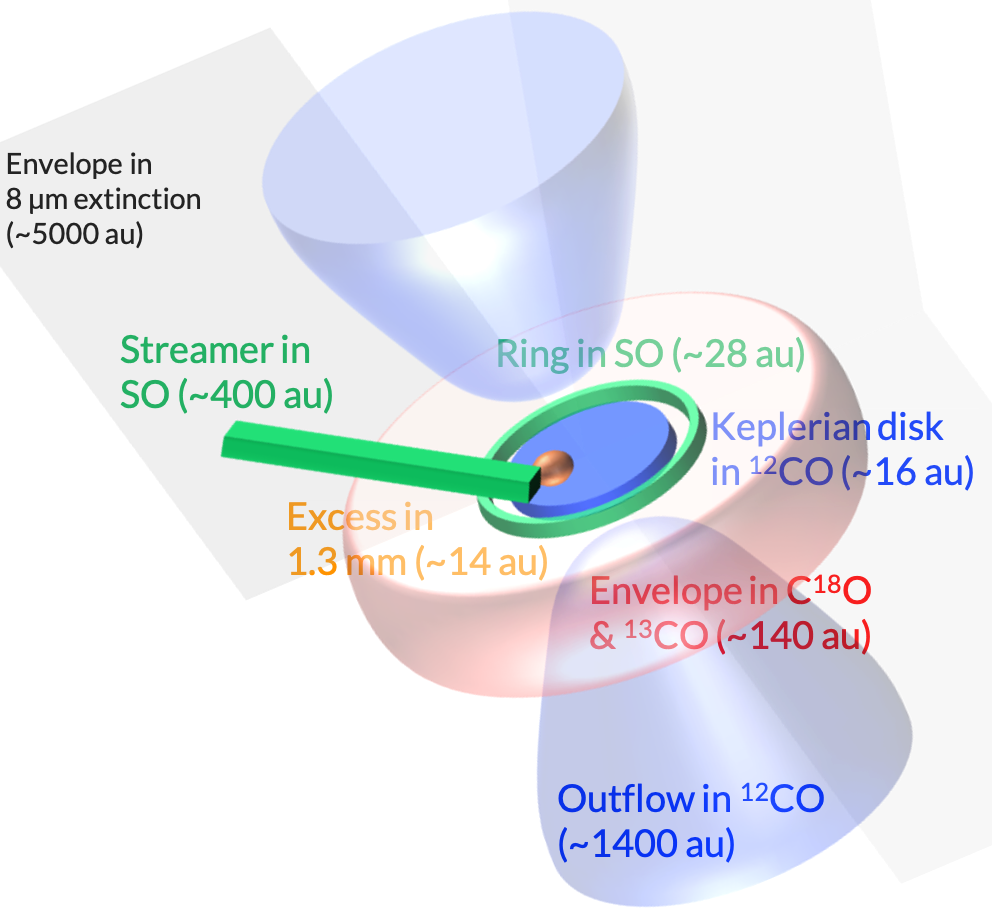}{0.5\textwidth}{}
\caption{Schematic picture of the Keplerian disk, ring, envelope, streamer, outflow identified in the CO isotopologue and SO lines, and the eastern excess in the 1.3 mm continuum emission in the Class 0 protostar IRAS16253-2429. The gray background represents the large-scale envelope identified in 8 $\micron$ extinction \citep{tobi10}. The radii or lengths for each of the structures are indicated as well in units of au.
\label{fig:schem}}
\end{figure}

\citet{tobi10} show the envelope of I16253 in the $8~\micron$ extinction observations at a $\sim 2\arcsec$ resolution using InfraRed Array Camera (IRAC) on the {\it Spitzer} telescope. The extinction is mainly concentrated in the $\sim 40\arcsec$-long outflow cavity wall, and the northeastern wall shows higher extinction than the other walls. This may suggest the existence of inhomogeneities in the ambient material that could produce the anisotropic streamer observed in our SO result. On the other hand, if the anisotropy in the SO line is due to an inhomogeneous density structure, similar anisotropy could also be seen in the C$^{18}$O and $^{13}$CO lines, unlike our results. This may suggest that the anisotropy in the SO line could be caused by other factors than density, such as temperature, shock, or chemistry. Figure \ref{fig:schem} summarizes the structures identified around I16253 in our ALMA observations, along with the large-scale envelope, denoting the radius or length of each structure.

\subsection{Mass Accretion and Other Quantities in I16253} \label{sec:acc}
The central stellar mass of I16253 has been directly estimated to be $0.12-0.17~\Msun$ by identifying the Keplerian rotation in its disk in this paper, which ranged from $\sim 0.02$ to $\sim 0.12~\Msun$ in previous works (Section \ref{sec:intro}). In contrast, some previous works estimated the central stellar mass of a protostar or a proto-BD from the mass accretion rate from the disk to the central object $\dot{M}_{\rm acc}$.
We discuss the uncertainties for the method using $\dot{M}_{\rm acc}$ along with the accretion rate derived from the updated stellar mass.

The central stellar mass of I16253 that we have estimated requires a mass accretion rate of $\dot{M}_{\rm acc}=L_{\rm bol}R_* / GM_* =(0.9-1.3)\times 10^{-7}~\Msun~{\rm yr}^{-1}$, where the bolometric luminosity is $L_{\rm bol}=0.16~\Lsun$, the stellar radius is assumed to be $R_*=3~\Rsun$, $G$ is the gravitational constant, and the stellar mass is $M_*=0.12-0.17~\Msun$.
The stellar radius is predicted in numerical simulations to be within $\sim 30\%$ of $3~\Rsun$ \citep[e.g.,][]{ma.in00, vo.ba15}.
In comparison, \citet{hsie16} estimated the mass accretion rate of I16253 from its outflow force.
The outflow force, $F_{\rm out}\sim 7\times 10^{-7}~\Msun~\kms~{\rm yr}^{-1}$, was measured through IRAM 30-m and APEX observations in $^{12}$CO lines at an angular resolution of $\sim 11\arcsec$. This force was converted to a mass accretion rate of
$\dot{M}_{\rm acc}\sim 5\times 10^{-7}~\Msun~{\rm yr}^{-1}$ (after the distance correction from 125 to 139 pc) as $\dot{M}_{\rm acc}=F_{\rm out} /\epsilon V_W f_{\rm ent}$, where the ratio between mass loss and accretion rates $\epsilon=\dot{M}_W/\dot{M}_{\rm acc}$ is assumed to be 0.1, the wind (jet) velocity $V_W$ is assumed to be 150~$\kms$, and the entrainment efficiency $f_{\rm ent}$ is assumed to be 0.1.
$F_{\rm out}$ is estimated from observations in the $^{12}$CO $J=2-1$ line using IRAM 30-m and $^{12}$CO $J=7-6$ and $J=6-5$ lines using APEX, with the optical depth correction and assumptions that the kinetic temperature is 40~K and the H$_2$ number density is $10^5~{\rm cm}^{-3}$. This mass accretion rate is $\sim 5$ times higher than the one derived from the updated stellar mass, whereas being similar to that of two proto-BD candidates, IC348-SMM2D and L328-IRS, $\sim 2-9\times 10^{-7}~\Msun~{\rm yr}^{-1}$.

This difference can be explained by large uncertainties in the conversion from the outflow force to the mass accretion rate. The mass loss and accretion ratio $\epsilon = \dot{M}_{\rm W}/\dot{M}_{\rm acc}$, the jet velocity $V_W$, and the efficiency $f_{\rm ent}$ can vary from $\sim 0.01$ to $\sim 0.5$ \citep{elle13,podi21}, $\sim 30$ to $\sim 160~\kms$ \citep[jets in SiO $5-4$;][]{podi21}, and 0.1 to 0.25 \citep{an.ba99}, respectively, in the protostellar phase. 
A theoretical model called X-wind suggests that $\epsilon$ and $V_W$ are anti-correlated and the factor of $\epsilon V_W$ varies only around $\sim 50$ to $\sim 70~\kms$ \citep{na.sh94} in the T Tauri phase. However, these velocities are $\gtrsim 4$ times higher than the typical velocity for protostars, $\sim 15~\kms$, as adopted in \citet{hsie16}, and the anti-correlation is not observationally confirmed in the protostellar phase.
Those large uncertainties imply that $\dot{M}_{\rm acc}$ calculated by \citet{hsie16} could be an order of magnitude lower. The abovementioned extreme values provide a range of $4\times 10 ^{-8}$ to $2\times 10^{-5}~\Msun ~{\rm yr}^{-1}$. For this reason, if the central stellar mass is estimated from the Keplerian rotation, the mass accretion rate could also be estimated better than from the outflow observations.

The updated mass accretion rate of I16253, $\sim 1\times 10^{-7}~\Msun~{\rm yr}^{-1}$, cannot provide $M_* =0.14~\Msun$ within the lifetime of Class 0 $\lesssim $0.2~Myr. In other words, the updated central stellar mass is large against the luminosity of I16253, as the required time can be estimated to be $M_*/\dot{M}_{\rm acc}=GM_*^2 / R_* L_{\rm bol}\sim 1$~Myr. This suggests that I16253 likely experienced an accretion burst (or stronger accretion) in the past. The presence of an accretion burst is supported by the episodic mass ejection as seen in the $^{12}$CO PV diagram along the outflow axis (Section \ref{sec:12co}). The Class 0 age is also long enough to experience bursts with a typical interval of 2400~years in the Class 0 phase \citep{hsie19b}.
Once an accretion burst occurs, the luminosity of a protostar increases resulting in higher temperatures in the surrounding envelope at a given radius. This causes CO molecules to sublimate from the icy grain mantles in an extended region of the envelope. After the protostar has returned to its quiescent state and the luminosity decreased again, the molecules remain in the gas-phase for a period of time before freezing out again. This freeze-out time-scale depends on the density at a given radius \citep{ro.ch03} but is typically of order $10^4$~years where CO sublimation due to an accretion burst occurs around Solar-type protostars \citep[e.g.,][]{jorg15}.
From this point of view, the size of the C$^{18}$O emission observed in I16253 (Figure \ref{fig:c18o13co}a), $\gtrsim 150$~au, also supports that an accretion burst happened because this size is much larger than expected from the current luminosity of this protostar: The C$^{18}$O emission radius is predicted to be $\sim 30$~au with the current luminosity of I16253, $0.16~\Lsun$ \citep{jorg15} \citep[see also][]{lee07}. 

Based on the measured $M_*$ and the above discussion about the mass accretion and the luminosity, we conclude that I16253 is not a proto-BD even though its luminosity and mass accretion rate are similar to those of the two proto-BD candidates.
The core (or envelope) mass of I16253 is estimated to be $\sim 1~\Msun$ from 1.1 mm observations with Bolocam on the CSO telescope at an angular resolution of $31\arcsec$ \citep{youn06} and $8~\micron$ observations with IRAC on the {\it Spitzer} telescope at an angular resolution of $2\arcsec$ \citep{tobi10}. If this envelope mass accretes onto I16253 with a typical star formation efficiency \citep[$30\% \pm 10\%$ calculated from the dense core mass function and initial mass function;][]{alve07}, a mass of 0.2 to 0.4~$\Msun$ will be added to the central stellar mass. This also supports an idea that I16253 will ultimately obtain mass high enough to fuse hydrogen (i.e., $M_* >0.08~\Msun$).

The Keplerian disk around I16253 has been kinematically identified for the first time. Without any clear identification of the Keplerian disk, previous works have attempted to estimate the central stellar mass using the infalling motion or the outflow force (through the mass accretion rate) in I16253 and suggested this protostar as a proto-BD candidate. Similarly, the central stellar mass of proto-BD candidates was estimated in those methods in some previous works (Section \ref{sec:intro}), without identifying a Keplerian disk.
Our study demonstrates that a proto-BD and a very low mass protostar must be identified through the dynamical central stellar mass derived from the Keplerian rotation of its disk, rather than its infall motion or outflow force.

The accurate mass estimation enables us to compare physical quantities in the I16253 system with scaling relations among young stellar objects. For example, its disk radius, $16-19$~au, appears consistent with or slightly lower than the scaling relation between the disk radius and the central stellar mass found with the protostellar sample of \citet{yen17}. Its mass accretion rate, $(0.9-1.3)\times 10^{-7}~\Msun~{\rm yr}^{-1}$, and disk mass, $\gtrsim 2\times 10^{-3}~\Msun$, are consistent with the scaling relation found with the Class I sample of \citet{fior22}. More observations around the BD mass threshold with accurate mass estimations will be necessary to determine whether these scaling relations hold down to the BD mass regime and bridge star formation and BD formation.

\section{Conclusions} \label{sec:conc}
As a part of the ALMA large project eDisk, we have observed the Class 0 protostar IRAS16253-2429, which has been suggested to be a proto-brown dwarf candidate in previous works, in the 1.3 mm continuum, $^{12}$CO $J=2-1$, C$^{18}$O $J=2-1$, $^{13}$CO $J=2-1$, SO $J_N = 6_5 - 5_4$, and other molecular lines at an angular resolution of $0\farcs 07$ ($\sim 10$~au). Our results provide a typical picture of protostars with a very low stellar mass close to the brown dwarf threshold (Figure \ref{fig:schem}). The main results are summarized below.

\begin{enumerate}
\item
The continuum emission shows structures from a $\sim 600$-au scale down to a $\sim 15$-au scale. The main component shows a disk-like structure with a radius of $\sim 20$~au. The emission is extended to the southeast along the major axis. These extensions can be interpreted as an enhancement due to a streamer from the east and the near-side (southwestern side) wall of the disk-like structure.
\item
The $^{12}$CO emission overall traces a clear bipolar outflow up to a $\sim 3000$-au scale. The outflow suggests episodic mass ejections. Furthermore, the $^{12}$CO emission on the midplane shows a velocity gradient along the disk's major axis, implying rotation of the disk. 
\item
We analyzed the $^{12}$CO major-axis position-velocity diagram by the edge and ridge methods and identified a Keplerian disk in IRAS16253-2429 for the first time. From this identification of the Keplerian rotation in both methods, the central stellar mass is estimated to be $0.12 - 0.17~\Msun$. This mass leads us to conclude that IRAS16253-2429 is unlikely a proto-brown dwarf but a very low mass protostar.
\item
The C$^{18}$O and $^{13}$CO emissions trace the infalling and rotating envelope as reported in previous observational works. The major- and minor-axis position-velocity diagrams in these lines are consistent with the UCM envelope model with the central stellar mass of $0.14~\Msun$ and the disk (centrifugal) radius of $r\sim 16$~au. Their moment 0 maps exhibit stronger emission intensities on the eastern peak, which could result from the streamer mentioned below.
\item
The SO emission shows a different velocity structure from the CO isotopologues. Its double peaks in the moment 0 map and linear velocity gradient in the major-axis position-velocity diagram suggest that this emission traces a ring ($r\sim 28$~au) due to the accretion shock between the disk and the envelope. This emission also shows a streamer from the eastern side, which can be explained with a ballistic, parabolic flow at $30\arcdeg$ above the midplane extracted from the UCM envelope model.
\end{enumerate}

\section*{Acknowledgements}
This paper makes use of the following ALMA data: ADS/JAO.ALMA\#2019.1.00261.L and ADS/JAO.ALMA\#2019.A.00034.S. ALMA is a partnership of ESO (representing its member states), NSF (USA) and NINS (Japan), together with NRC (Canada), MOST and ASIAA (Taiwan), and KASI (Republic of Korea), in cooperation with the Republic of Chile. The Joint ALMA Observatory is operated by ESO, AUI/NRAO and NAOJ. The National Radio Astronomy Observatory is a facility of the National Science Foundation operated
under cooperative agreement by Associated Universities, Inc. 
W.K. was supported by the National Research Foundation of Korea (NRF) grant funded by the Korea government (MSIT) (NRF-2021R1F1A1061794).
N.O. acknowledges support from National Science and Technology Council (NSTC) in Taiwan through the grants NSTC 109-2112-M-001-051 and 110-2112-M-001-031.
J.K.J. acknowledge support from the Independent Research Fund Denmark (grant No. 0135-00123B).
J.J.T. acknowledges support from NASA XRP 80NSSC22K1159. N.O. acknowledges support from National Science and Technology Council (NSTC) in Taiwan through the grants NSTC 109-2112-M-001-051 and 110-2112-M-001-031.
Y.A. acknowledges support by NAOJ ALMA Scientific Research Grant code 2019-13B, Grant-in-Aid for Scientific Research (S) 18H05222, and Grant-in-Aid for Transformative Research Areas (A) 20H05844 and 20H05847.
IdG acknowledges support from grant PID2020-114461GB-I00, funded by MCIN/AEI/10.13039/501100011033.
PMK acknowledges support from NSTC 108-2112- M-001-012, NSTC 109-2112-M-001-022 and NSTC 110-2112-M-001-057.
SPL and TJT acknowledge grants from the National Science and Technology Council of Taiwan 106-2119-M-007-021-MY3 and 109-2112-M-007-010-MY3. J.K.J. acknowledges support from the Independent Research Fund Denmark (grant No. 0135-00123B).
C.W.L. is supported by the Basic Science Research Program through the National Research Foundation of Korea (NRF) funded by the Ministry of Education, Science and Technology (NRF- 2019R1A2C1010851), and by the Korea Astronomy and Space Science Institute grant funded by the Korea government (MSIT; Project No. 2022-1-840-05). 
JEL was supported by the National Research Foundation of Korea (NRF) grant funded by the Korean government (MSIT) (grant number 2021R1A2C1011718).
ZYL is supported in part by NASA 80NSSC20K0533 and NSF AST-1910106.
ZYDL acknowledges support from NASA 80NSSC18K1095, the Jefferson Scholars Foundation, the NRAO ALMA Student Observing Support (SOS) SOSPA8-003, the Achievements Rewards for College Scientists (ARCS) Foundation Washington Chapter, the Virginia Space Grant Consortium (VSGC), and UVA research computing (RIVANNA).
LWL acknowledges support from NSF AST-2108794.
S.N. acknowledges support from the National Science Foundation through the Graduate Research Fellowship Program under Grant No. 2236415. Any opinions, findings, and conclusions or recommendations expressed in this material are those of the authors and do not necessarily reflect the views of the National Science Foundation.
R.S. acknowledge support from the Independent Research Fund Denmark (grant No. 0135-00123B).
S.T. is supported by JSPS KAKENHI grant Nos. 21H00048 and 21H04495, and by NAOJ ALMA Scientific Research grant No. 2022-20A.
JPW acknowledges support from NSF AST-2107841.
H.-W.Y. acknowledges support from the National Science and Technology Council (NSTC) in Taiwan through the grant NSTC 110-2628-M-001-003-MY3 and from the Academia Sinica Career Development Award (AS-CDA-111-M03).
%

%

\vspace{5mm}
\facilities{ALMA}


\software{astropy \citep{astr13, astr18}, CASA \citep{mcmu07}, eDisk script \citep{tobi23}, ptemcee \citep{fore13, vous16}, SLAM \citep{as.sa23}}



\appendix

\section{Other molecular lines} \label{sec:otherlines}

\begin{figure}[htbp]
\gridline{
\fig{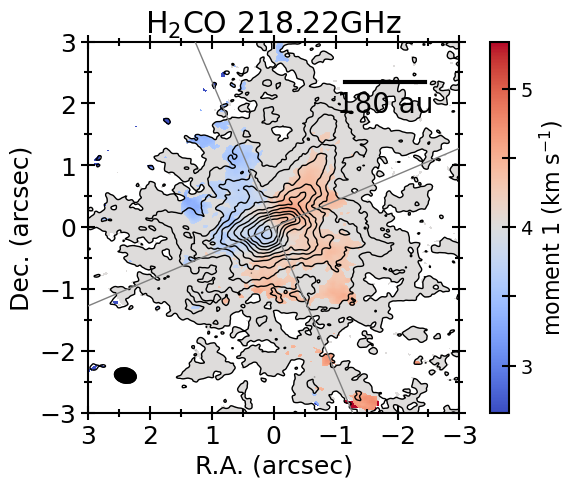}{0.33\textwidth}{(a)}
\fig{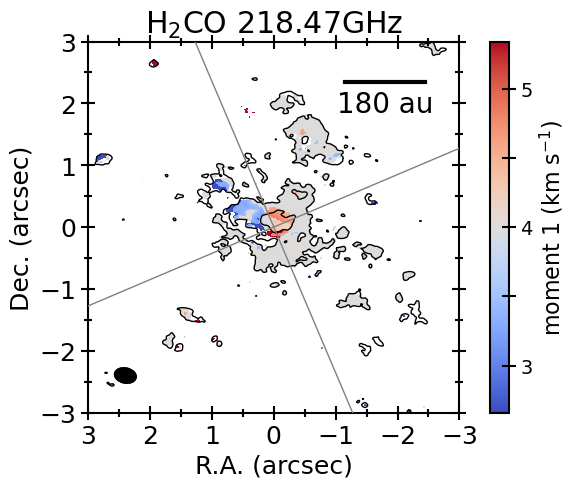}{0.33\textwidth}{(b)}
\fig{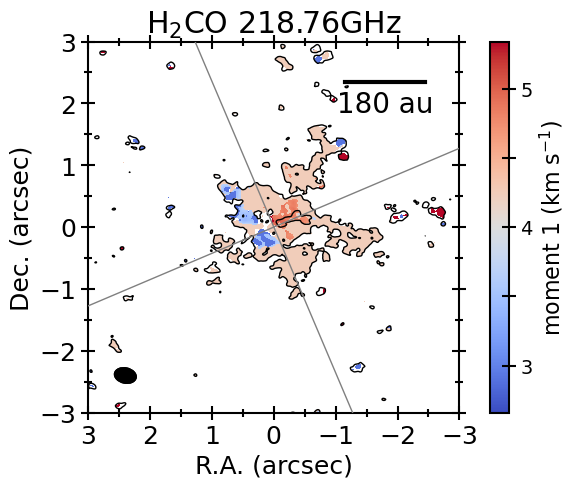}{0.33\textwidth}{(c)}
}
\gridline{
\fig{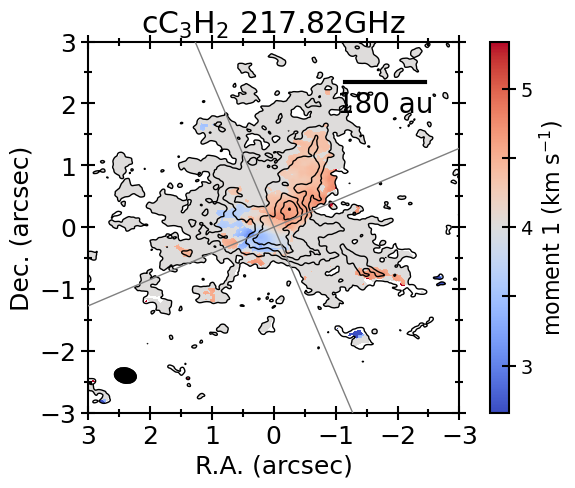}{0.33\textwidth}{(d)}
\fig{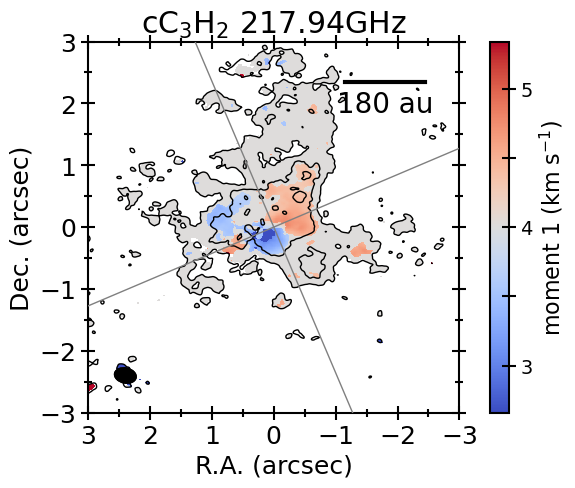}{0.33\textwidth}{(e)}
\fig{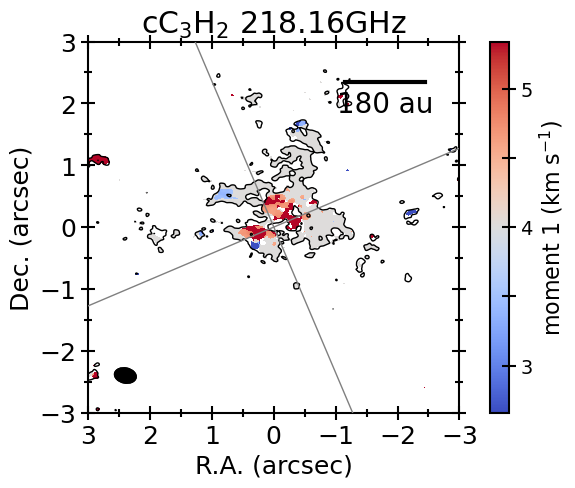}{0.33\textwidth}{(f)}
}
\gridline{
\fig{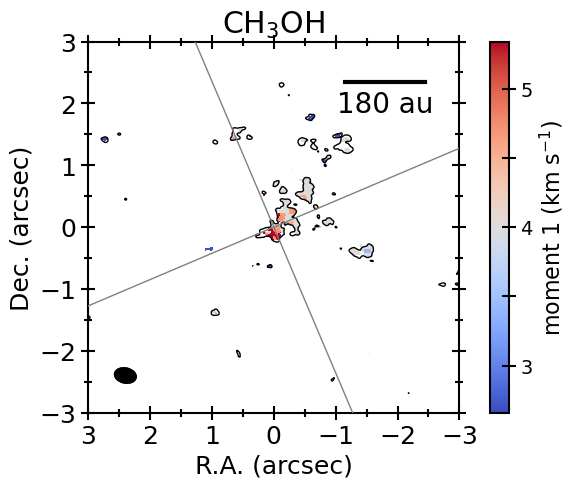}{0.33\textwidth}{(g)}
\fig{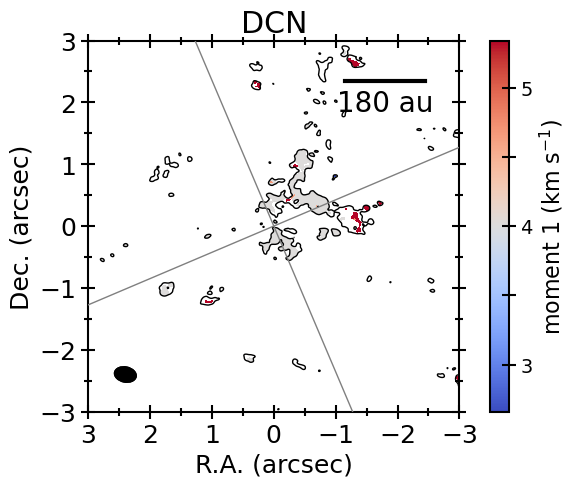}{0.33\textwidth}{(h)}
\fig{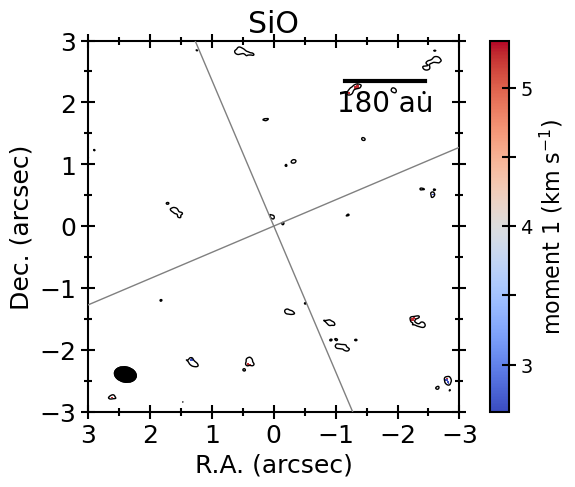}{0.33\textwidth}{(i)}
}
\caption{Moment 0 and 1 maps in 5 other molecular lines (9 transitions) observed in the ALMA observations of the eDisk project. These images are made with the robust parameter of 2. The emission is integrated from 2.66 to $5.34~\kms$ (3 channels) for all lines. The contour levels are in $3\sigma$ steps from $3\sigma$, where $1\sigma$ is $1.5~\mJB~\kms$. The ellipse at the lower left corner in each panel denotes the synthesized beam, $\sim 0\farcs 38\times 0\farcs 27\ (78\arcdeg)$.
The diagonal lines are the major and minor axes of the continuum emission, P.A.$=113\arcdeg$ and $23\arcdeg$.
\label{fig:others}}
\end{figure}

Figure \ref{fig:others} shows the moment 0 and 1 maps of the lines observed in the eDisk project toward I16253, except for the $^{12}$CO, C$^{18}$O, $^{13}$CO, and SO lines. These lines were observed from the wide spectral windows otherwise intended for continuum measurements. The three H$_2$CO lines have different upper state energies, depending on the rest frequencies (10.5 K; 218.22 GHz, 57.6 K; 218.47 GHz, 57.6 K; 218.76 GHz). The three cyclic C$_3$H$_2$ lines also have different upper state energies (28.2 K; 217.82 GHz, 25.0 K; 217.94 GHz, 24.9 K; 218.16 GHz). The CH$_3$OH, DCN, and SiO lines have upper state energies of 35.0, 10.4 and 20.8~K, respectively. In comparison, the $^{12}$CO, C$^{18}$O, $^{13}$CO, and SO lines have upper state energies of 5.5, 5.3, 5.3, and 24.4~K, respectively. All the panels in Figure \ref{fig:others} are made from three channels centered at the systemic velocity ($V_{\rm LSR}=2.66 - 5.34~\kms$) and show the same spatial and velocity ranges. 

\begin{thebibliography}{}
\bibitem[Alves et al.(2017)]{alve17} Alves, F.~O., Girart, J.~M., Caselli, P., et al.\ 2017, \aap, 603, L3. doi:10.1051/0004-6361/201731077
\bibitem[Alves et al.(2007)]{alve07} Alves, J., Lombardi, M., \& Lada, C.~J.\ 2007, \aap, 462, L17. doi:10.1051/0004-6361:20066389
\bibitem[Andr{\'e} et al.(1999)]{an.ba99} Andr{\'e}, P., Motte, F., \& Bacmann, A.\ 1999, \apjl, 513, L57. doi:10.1086/311908
\bibitem[Andrews \& Williams(2005)]{an.wi05} Andrews, S.~M. \& Williams, J.~P.\ 2005, \apj, 631, 1134. doi:10.1086/432712
\bibitem[Aso \& Machida(2020)]{as.ma20} Aso, Y. \& Machida, M.~N.\ 2020, \apj, 905, 174. doi:10.3847/1538-4357/abc6fc
\bibitem[Aso \& Sai(2023)]{as.sa23} Aso, Y. \& Sai, J.\ 2023, jinshisai/SLAM: First Release of SLAM, v1.0.0, Zenodo, doi:10.5281/zenodo.7783868
\bibitem[Aso et al.(2017)]{aso17} Aso, Y., Ohashi, N., Aikawa, Y., et al.\ 2017, \apj, 849, 56. doi:10.3847/1538-4357/aa8264
\bibitem[Aso et al.(2015)]{aso15} Aso, Y., Ohashi, N., Saigo, K., et al.\ 2015, \apj, 812, 27. doi:10.1088/0004-637X/812/1/27
\bibitem[Aso et al.(2021)]{aso21} Aso, Y., Kwon, W., Hirano, N., et al.\ 2021, \apj, 920, 71. doi:10.3847/1538-4357/ac15f3
\bibitem[Astropy Collaboration et al.(2018)]{astr18} Astropy Collaboration, Price-Whelan, A.~M., Sip{\H{o}}cz, B.~M., et al.\ 2018, \aj, 156, 123. doi:10.3847/1538-3881/aabc4f
\bibitem[Astropy Collaboration et al.(2013)]{astr13} Astropy Collaboration, Robitaille, T.~P., Tollerud, E.~J., et al.\ 2013, \aap, 558, A33. doi:10.1051/0004-6361/201322068
\bibitem[Basu et al.(2012)]{ba.vo12} Basu, S., Vorobyov, E.~I., \& DeSouza, A.~L.\ 2012, First Stars IV - from Hayashi to the Future -, 1480, 63. doi:10.1063/1.4754329
\bibitem[Bate(2012)]{bate12} Bate, M.~R.\ 2012, \mnras, 419, 3115. doi:10.1111/j.1365-2966.2011.19955.x
\bibitem[Beckwith et al.(1990)]{beck90} Beckwith, S.~V.~W., Sargent, A.~I., Chini, R.~S., et al.\ 1990, \aj, 99, 924. doi:10.1086/115385
\bibitem[Cassen \& Moosman(1981)]{ca.mo81} Cassen, P. \& Moosman, A.\ 1981, \icarus, 48, 353. doi:10.1016/0019-1035(81)90051-8
\bibitem[Chabrier(2002)]{chab02} Chabrier, G.\ 2002, \apj, 567, 304. doi:10.1086/324716
\bibitem[Cabrit(1989)]{cabr89} Cabrit, S.\ 1989, European Southern Observatory Conference and Workshop Proceedings, 33, 119
\bibitem[de Geus et al.(1989)]{dege89} de Geus, E.~J., de Zeeuw, P.~T., \& Lub, J.\ 1989, \aap, 216, 44
\bibitem[de Gregorio-Monsalvo et al.(2016)]{degre16} de Gregorio-Monsalvo, I., Barrado, D., Bouy, H., et al.\ 2016, \aap, 590, A79. doi:10.1051/0004-6361/201424149
\bibitem[Dunham et al.(2008)]{dunh08} Dunham, M.~M., Crapsi, A., Evans, N.~J., et al.\ 2008, \apjs, 179, 249. doi:10.1086/591085
\bibitem[Ellerbroek et al.(2013)]{elle13} Ellerbroek, L.~E., Podio, L., Kaper, L., et al.\ 2013, \aap, 551, A5. doi:10.1051/0004-6361/201220635
\bibitem[Fiorellino et al.(2022)]{fior22} Fiorellino, E., Tychoniec, {\L}., Manara, C.~F., et al.\ 2022, \apjl, 937, L9. doi:10.3847/2041-8213/ac8fee
\bibitem[Foreman-Mackey et al.(2013)]{fore13} Foreman-Mackey, D., Hogg, D.~W., Lang, D., et al.\ 2013, \pasp, 125, 306. doi:10.1086/670067
\bibitem[Garufi et al.(2022)]{garu22} Garufi, A., Podio, L., Codella, C., et al.\ 2022, \aap, 658, A104. doi:10.1051/0004-6361/202141264
\bibitem[Hanawa et al.(2022)]{hana22} Hanawa, T., Sakai, N., \& Yamamoto, S.\ 2022, \apj, 932, 122. doi:10.3847/1538-4357/ac6e6a
\bibitem[Hsieh et al.(2019)]{hsie19} Hsieh, T.-H., Hirano, N., Belloche, A., et al.\ 2019, \apj, 871, 100. doi:10.3847/1538-4357/aaf4fe
\bibitem[Hsieh et al.(2019)]{hsie19b} Hsieh, T.-H., Murillo, N.~M., Belloche, A., et al.\ 2019, \apj, 884, 149. doi:10.3847/1538-4357/ab425a
\bibitem[Hsieh et al.(2016)]{hsie16} Hsieh, T.-H., Lai, S.-P., Belloche, A., et al.\ 2016, \apj, 826, 68. doi:10.3847/0004-637X/826/1/68
\bibitem[Hu{\'e}lamo et al.(2017)]{huel17} Hu{\'e}lamo, N., de Gregorio-Monsalvo, I., Palau, A., et al.\ 2017, \aap, 597, A17. doi:10.1051/0004-6361/201628510
\bibitem[J{\o}rgensen et al.(2015)]{jorg15} J{\o}rgensen, J.~K., Visser, R., Williams, J.~P., et al.\ 2015, \aap, 579, A23. doi:10.1051/0004-6361/201425317
\bibitem[Kido et al.(2023)]{kido23} Kido, M., Takakuwa, S., Saigo, K., et al.\ 2023, \apj, 953, 190. doi:10.3847/1538-4357/acdd7a
\bibitem[Kim et al.(2019)]{kim19} Kim, G., Lee, C.~W., Maheswar, G., et al.\ 2019, \apjs, 240, 18. doi:10.3847/1538-4365/aaf889
\bibitem[Knude \& Hog(1998)]{kn.ho98} Knude, J. \& Hog, E.\ 1998, \aap, 338, 897
\bibitem[Lee(2020)]{lee20} Lee, C.-F.\ 2020, \aapr, 28, 1. doi:10.1007/s00159-020-0123-7
\bibitem[Lee et al.(2018)]{lee18} Lee, C.~W., Kim, G., Myers, P.~C., et al.\ 2018, \apj, 865, 131. doi:10.3847/1538-4357/aadcf6
\bibitem[Lee(2007)]{lee07} Lee, J.-E.\ 2007, Journal of Korean Astronomical Society, 40, 83. doi:10.5303/JKAS.2007.40.4.083
\bibitem[Machida et al.(2009)]{mach09} Machida, M.~N., Inutsuka, S.-. ichiro ., \& Matsumoto, T.\ 2009, \apjl, 699, L157. doi:10.1088/0004-637X/699/2/L157
\bibitem[Machida et al.(2010)]{mach10} Machida, M.~N., Inutsuka, S.-. ichiro ., \& Matsumoto, T.\ 2010, \apj, 724, 1006. doi:10.1088/0004-637X/724/2/1006
\bibitem[Machida et al.(2014)]{mach14} Machida, M.~N., Inutsuka, S.-. ichiro ., \& Matsumoto, T.\ 2014, \mnras, 438, 2278. doi:10.1093/mnras/stt2343
\bibitem[Maret et al.(2020)]{mare20} Maret, S., Maury, A.~J., Belloche, A., et al.\ 2020, \aap, 635, A15. doi:10.1051/0004-6361/201936798
\bibitem[Masunaga \& Inutsuka(2000)]{ma.in00} Masunaga, H. \& Inutsuka, S.-. ichiro .\ 2000, \apj, 531, 350. doi:10.1086/308439
\bibitem[McMullin et al.(2007)]{mcmu07} McMullin, J.~P., Waters, B., Schiebel, D., et al.\ 2007, Astronomical Data Analysis Software and Systems XVI, 376, 127
\bibitem[Najita \& Shu(1994)]{na.sh94} Najita, J.~R. \& Shu, F.~H.\ 1994, \apj, 429, 808. doi:10.1086/174365
\bibitem[Ohashi et al.(2014)]{ohas14} Ohashi, N., Saigo, K., Aso, Y., et al.\ 2014, \apj, 796, 131. doi:10.1088/0004-637X/796/2/131
\bibitem[Ohashi et al.(2023)]{ohas23} Ohashi, N., Tobin, J.~J., J{\o}rgensen, J.~K., et al.\ 2023, \apj, 951, 8. doi:10.3847/1538-4357/acd384
\bibitem[Padoan \& Nordlund(2004)]{pa.no04} Padoan, P. \& Nordlund, {\r{A}}.\ 2004, \apj, 617, 559. doi:10.1086/345413
\bibitem[Palau et al.(2014)]{pala14} Palau, A., Zapata, L.~A., Rodr{\'\i}guez, L.~F., et al.\ 2014, \mnras, 444, 833. doi:10.1093/mnras/stu1461
\bibitem[Park et al.(2021)]{park21} Park, W., Lee, J.-E., Contreras Pe{\~n}a, C., et al.\ 2021, \apj, 920, 132. doi:10.3847/1538-4357/ac1745
\bibitem[Pineda et al.(2020)]{pine20} Pineda, J.~E., Segura-Cox, D., Caselli, P., et al.\ 2020, Nature Astronomy, 4, 1158. doi:10.1038/s41550-020-1150-z
\bibitem[Podio et al.(2021)]{podi21} Podio, L., Tabone, B., Codella, C., et al.\ 2021, \aap, 648, A45. doi:10.1051/0004-6361/202038429
\bibitem[Rodgers \& Charnley(2003)]{ro.ch03} Rodgers, S.~D. \& Charnley, S.~B.\ 2003, \apj, 585, 355. doi:10.1086/345497
\bibitem[Ru{\'\i}z-Rodr{\'\i}guez et al.(2022)]{ruiz22} Ru{\'\i}z-Rodr{\'\i}guez, D.~A., Williams, J.~P., Kastner, J.~H., et al.\ 2022, \mnras, 515, 2646. doi:10.1093/mnras/stac1879
\bibitem[Sai et al.(2022)]{sai22} Sai, J., Ohashi, N., Maury, A.~J., et al.\ 2022, \apj, 925, 12. doi:10.3847/1538-4357/ac341d
\bibitem[Sai et al.(2020)]{sai20} Sai, J., Ohashi, N., Saigo, K., et al.\ 2020, \apj, 893, 51. doi:10.3847/1538-4357/ab8065
\bibitem[Sakai et al.(2016)]{saka16} Sakai, N., Oya, Y., L{\'o}pez-Sepulcre, A., et al.\ 2016, \apjl, 820, L34. doi:10.3847/2041-8205/820/2/L34
\bibitem[Santamar{\'\i}a-Miranda et al.(2021)]{sant21} Santamar{\'\i}a-Miranda, A., de Gregorio-Monsalvo, I., Plunkett, A.~L., et al.\ 2021, \aap, 646, A10. doi:10.1051/0004-6361/202039419
\bibitem[Seifried et al.(2016)]{seif16} Seifried, D., S{\'a}nchez-Monge, {\'A}., Walch, S., et al.\ 2016, \mnras, 459, 1892. doi:10.1093/mnras/stw785
\bibitem[Shariff et al.(2022)]{shar22} Shariff, K., Gorti, U., \& Melon Fuksman, J.~D.\ 2022, \mnras, 514, 5548. doi:10.1093/mnras/stac1186
\bibitem[Stamatellos \& Whitworth(2009)]{st.wh09} Stamatellos, D. \& Whitworth, A.~P.\ 2009, 15th Cambridge Workshop on Cool Stars, Stellar Systems, and the Sun, 1094, 557. doi:10.1063/1.3099172
\bibitem[Thieme et al.(2022)]{thie22} Thieme, T.~J., Lai, S.-P., Lin, S.-J., et al.\ 2022, \apj, 925, 32. doi:10.3847/1538-4357/ac382b
\bibitem[Tobin et al.(2023)]{tobi23} Tobin, J.\ 2023, eDisk data reduction scripts, 1.0.0, Zenodo, doi:10.5281/zenodo.7986682
\bibitem[Tobin et al.(2012)]{tobi12} Tobin, J.~J., Hartmann, L., Bergin, E., et al.\ 2012, \apj, 748, 16. doi:10.1088/0004-637X/748/1/16
\bibitem[Tobin et al.(2010)]{tobi10} Tobin, J.~J., Hartmann, L., Looney, L.~W., et al.\ 2010, \apj, 712, 1010. doi:10.1088/0004-637X/712/2/1010
\bibitem[Tobin et al.(2020)]{tobi20} Tobin, J.~J., Sheehan, P.~D., Megeath, S.~T., et al.\ 2020, \apj, 890, 130. doi:10.3847/1538-4357/ab6f64
\bibitem[Ulrich(1976)]{ulri76} Ulrich, R.~K.\ 1976, \apj, 210, 377. doi:10.1086/154840
\bibitem[Vorobyov \& Basu(2015)]{vo.ba15} Vorobyov, E.~I. \& Basu, S.\ 2015, \apj, 805, 115. doi:10.1088/0004-637X/805/2/115
\bibitem[Vousden et al.(2016)]{vous16} Vousden, W.~D., Farr, W.~M., \& Mandel, I.\ 2016, \mnras, 455, 1919. doi:10.1093/mnras/stv2422
\bibitem[Whitworth \& Zinnecker(2004)]{wh.zi04} Whitworth, A.~P. \& Zinnecker, H.\ 2004, \aap, 427, 299. doi:10.1051/0004-6361:20041131
\bibitem[Wilner \& Welch(1994)]{wi.we94} Wilner, D.~J., \& Welch, W.~J.\ 1994, \apj, 427, 898 
\bibitem[Yen et al.(2019)]{yen19} Yen, H.-W., Gu, P.-G., Hirano, N., et al.\ 2019, \apj, 880, 69. doi:10.3847/1538-4357/ab29f8
\bibitem[Yen et al.(2017)]{yen17} Yen, H.-W., Koch, P.~M., Takakuwa, S., et al.\ 2017, \apj, 834, 178. doi:10.3847/1538-4357/834/2/178
\bibitem[Yen et al.(2014)]{yen14} Yen, H.-W., Takakuwa, S., Ohashi, N., et al.\ 2014, \apj, 793, 1. doi:10.1088/0004-637X/793/1/1
\bibitem[Yen et al.(2013)]{yen13} Yen, H.-W., Takakuwa, S., Ohashi, N., et al.\ 2013, \apj, 772, 22. doi:10.1088/0004-637X/772/1/22
\bibitem[Young et al.(2006)]{youn06} Young, K.~E., Enoch, M.~L., Evans, N.~J., et al.\ 2006, \apj, 644, 326. doi:10.1086/503327
\bibitem[Zucker et al.(2020)]{zuck20} Zucker, C., Speagle, J.~S., Schlafly, E.~F., et al.\ 2020, \aap, 633, A51. doi:10.1051/0004-6361/201936145
\bibitem[Zucker et al.(2019)]{zuck19} Zucker, C., Speagle, J.~S., Schlafly, E.~F., et al.\ 2019, \apj, 879, 125. doi:10.3847/1538-4357/ab2388

\end{thebibliography}



\end{document}